\documentclass[lettersize,journal]{IEEEtran}
\usepackage{amsmath,amssymb,amsfonts,hyperref}

\usepackage{array}
\usepackage{textcomp}
\usepackage{stfloats}
\usepackage{url}
\usepackage{verbatim}
\usepackage{graphicx}
\usepackage{cite}
\usepackage[normalem]{ulem}

\usepackage{multirow}
\usepackage{subcaption}
\usepackage{cleveref}
\usepackage[ruled,vlined]{algorithm2e}
\usepackage{setspace}
\usepackage{float}
\usepackage[]{caption}
\usepackage{tabularx}
\usepackage{booktabs}
\usepackage[font=small]{caption}
\usepackage{threeparttable}

\begin{document}

\title{EEG-BBNet: a Hybrid Framework for Brain Biometric using Graph Connectivity}

\author{Payongkit Lakhan, Nannapas Banluesombatkul, Natchaya Sricom, Korn Surapat, Ratha Rotruchiphong, Phattarapong Sawangjai, Tohru Yagi, Tulaya Limpiti$^{*}$ and Theerawit Wilaiprasitporn$^{*}$, \IEEEmembership{Member, IEEE}

\thanks{This work was supported by PTT Public Company Limited, The SCB Public Company Limited, The Office of the Permanent Secretary of the Ministry of Higher Education, Science, Research and Innovation, Thailand (RGNS63-252) and National Research Council of Thailand (N41A640131)  \textit{($^{*}$co-corresponding authors: Theerawit Wilaiprasitporn and Tulaya Limpiti).}}
\thanks{P. Lakhan, N. Banluesombatkul, N. Sricom and T. Wilaiprasitporn are with Bio-inspired Robotics and Neural Engineering (BRAIN) Lab, School of Information Science and Technology (IST), Vidyasirimedhi Institute of Science \& Technology (VISTEC), Rayong, Thailand (e-mail: theerawit.w@vistec.ac.th).}
\thanks{T. Limpiti are with the School of Engineering, King Mongkut's Institute of Technology Ladkrabang, Thailand. (e-mail:tulaya.li@kmitl.ac.th)}
\thanks{K. Surapat and R. Rotruchiphong are with Kamnoetvidya Science Academy School, Rayong, Thailand.}
\thanks{T. Yagi are with Yagi Lab, Department of Mechanical Engineering, Tokyo Institute of Technology, Tokyo, Japan.}
}

\markboth{Journal of \LaTeX\ Class Files,~Vol.~14, No.~8, August~2021}%
{P.Lakhan \MakeLowercase{\textit{et al.}}: EEG-BBNet: a Hybrid Framework for Brain Biometric using Graph Connectivity}


\maketitle

\begin{abstract}
% to be edited after everything else....

Brain biometrics based on electroencephalography (EEG) have been used increasingly for personal identification. Traditional machine learning techniques as well as modern day deep learning methods have been applied with promising results.  In this paper we present EEG-BBNet, a hybrid network which integrates convolutional neural networks (CNN) with graph convolutional neural networks (GCNN). The benefit of the CNN in automatic feature extraction and the capability of GCNN in learning connectivity between EEG electrodes through graph representation are jointly exploited. We examine various connectivity measures, namely the Euclidean distance, Pearson's correlation coefficient, phase-locked value, phase-lag index, and Rho index. The performance of the proposed method is assessed on a benchmark dataset consisting of various brain-computer interface (BCI) tasks and compared to other state-of-the-art approaches. We found that our models outperform all baselines in the event-related potential (ERP) task with an average correct recognition rates up to 99.26\% using intra-session data. EEG-BBNet with Pearson's correlation and RHO index provide the best classification results. In addition, our model demonstrates greater adaptability using inter-session and inter-task data.   
We also investigate the practicality of our proposed model with smaller number of electrodes. Electrode placements over the frontal lobe region appears to be most appropriate with minimal lost in performance.

\end{abstract}


\section{Introduction}
\label{sec:introduction}

\IEEEPARstart{A}{\MakeLowercase{uthentication}} has now been ingrained in our way of life.
The need to authenticate has grown exponentially over the past decades to the point where it is embedded in practically every digital object we own. %Even entry to a specific location, now requires identification to gain access. 
Due to the ease with which passwords and identification cards can be falsified, forgotten, or left behind, security systems have moved toward biometric-based personal identification and authentication using face, iris, fingerprint, or even voice scan \cite{unar2014review}. The primary advantages are their uniqueness, permanence, and universality.
Despite the increased security in the biometric features mentioned above, they remain vulnerable to spoofing attacks \cite{marcel2014handbook}.
Faking the fingerprint or eyeball, bypassing the face scan with photos and videos, hacking voice recognition with audio, or even forcing the user to unlock it directly are all possibilities. 
% As a result, the invader gains access to other people's resources. 
Electroencephalogram (EEG)-based authentication, i.e., \textit{brain biometrics}, is one of the solutions.
% By placing a set of electrodes over a human's scalp, it records electrical signals generated by neural activity in the human brain. 
This approach has established itself as a reliable personal identification method due to its variation in each individual \cite{palaniappan2007biometrics}. Additionally, it is significantly more difficult to forge than other biometrics and requires the user to be conscious and in a normal mental state to use.

%%%%

The EEG signals can be utilized to identify individuals by monitoring their brain activities when they are engaging in certain tasks or receiving external stimulations. For instance, resting EEGs can be obtained when the person is doing nothing and attempting to relax \cite{thomas2018eeg}.  Other brain responses can be acquired while the subjects are exposed to visual and auditory stimuli \cite{chen2016high}, imagining to move their limbs, or actually moving them \cite{patel2017biometrics}. Typically, the personal identification process involves two steps: extracting useful features from the acquired signal and classifying them using traditional machine learning techniques. Feature extraction can be done using various signal processing approaches, e.g., Power Spectral Density (PSD) \cite{thomas2018eeg}, Autoregressive model (AR) \cite{maiorana2016eeg}. Then, the features are fed into a classifier such as k-Nearest Neighbors (k-NN) \cite{vinothkumar2018task}, Linear Discriminant Analysis (LDA) \cite{lee2013study}, or artificial neural networks\cite{nieves2016automatic} to identify subjects. In recent years, Deep Learning (DL) have steadily gained popolarity as the processor of choice for artificial intelligence applications, including brain-based biometric systems  \cite{Gui2019ASO}. It combines feature extraction and classification steps into a single network, primarily over Convolutional Neural Networks (CNN). A variant of the CNN approach called Graph Convolutional Neural Networks (GCNN) has been extensively used in the last decade to understand social networks, computer vision, natural language processing, and life sciences \cite{wu2020comprehensive}. Despite its immense capabilities, several factors could impact the performance of GCNN and limit its usage in practice, for example, the performances when the model attempts to classify a newly seen data but is trained using data collected on different days or even different tasks. Furthermore, when the input graphs are constructed using connectivity between the EEG channels, the raw signals are usually discarded; even though the raw data itself contains information that is useful for personal identification, as demonstrated in prior studies.
%However, both traditional feature extraction and classification methods and classifiers contain many drawbacks which will be discussed in \Cref{sec:relatedWork}.
%%%%%

In this paper, we expand the capability of earlier brain biometric systems. We present \textit{EEG-BBNet}, a deep learning framework for brain biometric that combines the benefits of CNN and GCNN. We employ the CNN to automatically extract features from raw EEG data. 
The GCNN captures connectivity information between electrodes in terms of graph representation. The integration of CNN and GCNN cultivates more complete information from the available data. The performance of the proposed network and its practicality are assessed using datasets recorded on different days (sessions) and diverse BCI tasks. Furthermore, we explore the impacts of EEG electrode numbers and placements on the performance by employing the data pertaining to various cortical areas and the data conforming to consumer-grade EEG devices.

%  In this work, we propose a novel method for brain bio-metric using Graph Convolutional Neural Networks (GCN). The learning model is based on relational connectivity, representing EEG signals using graphs based on a neuro-theoretic perspective. A comprehensive study is conducted using a dataset with three recognition tasks and various sessions. The main contribution can be summarized as follows

% \begin{itemize}
%     \item We propose a novel layer-wise Graph Convolution Neural Networks-based deep learning framework for brain bio-metric which uses distance-based and correlation-based adjacency matrix.
%     \item We perform the graph-EEG analysis based on EEG electrode placement. With neuro-theoretic perspective, we divide the brain scalp into several regions to explore their differences.
%     \item We conduct Brain biometric model cross-session and cross-recognition tasks experiments.
% \end{itemize}

\section{Related Works}
\label{sec:relatedWork}
This section outlines previous works on brain biometric systems for personal identification. A brief summary of the algorithms mentioned herein are listed in \autoref{table_related_work} with specifications of the studied datasets including numbers of subject, numbers of channel, tasks, as well as input features and classifier choices.  
%The common evaluation metrics include the Correct Recognition Rate (CRR), False Acceptance Rate (FAR), False Rejection Rate (FRR), and Equal Error Rate (EER) \cite{Gui2019ASO}. 

% In this section, we review the development of EEG-based person identification. Because the purpose of this paper is to enhance the capability of and propose a learning model for EEG-based identification methods, the existing literature is reviewed under two headings which consisted of Machine Learning (ML) approaches and GCN approaches.

%=========================================================
\begin{table*}

\centering
\scriptsize
\begin{threeparttable}
\caption{Previous brain biometric studies.} 

\begin{tabular}{c c c c  p{0.01cm} c c  }
\toprule
% & \multicolumn{3}{c}{\textbf{Dataset}}  & &\multicolumn{2}{c}{\textbf{Method}} \\
\textbf{Study} & \textbf{No. of Subject} &	\textbf{No. of Channel} & \textbf{Task} && \textbf{Feature} & \textbf{Method} \\
 %& \textbf{Subjects} &	\textbf{Channels} & &&  &  \\% & \textbf{Performance} \\
\midrule 
 \cite{cai2015local} &40/33/11  &1	&   EC      &&	AR, PSD, WPT& k-NN/LLMNN	        \\% & CRR: 50.99\%-93.94\% /CRR: 58.47\%- 98.03\% \\

\cite{la2014human} &108	    &56 &   EC, EO  &&  PSD         & k-NN\\%Mahalanobis Similarity \\% & CRR: 75.86\%-90.49\% \\
 %\cite{ruiz2016cerebre}&50	    &1	&   ERP		&&  Raw         & KNN\\%    \\% & CRR: 60\%-86\% \\
 \cite{kumar2017bio} &50	    &14 &	MI	&& DFT/mean/std&	SVM	                \\% &FAR: 18\%-59\% \\
\cite{chu2017individual} & 120	    &64 &	EC, EO, ERP	&& Raw &	RF                \\% &FAR: 18\%-59\% \\

 \cite{el2018convolution} &4/10      &8	&   SSVEP   &&	Raw &	CNN	 \\% & CRR= 97.60 / 95.90 \% \\
 
\cite{sun2019eeg} &109	    & 4/16/32/64&	MI&&	Raw &	LSTM \\% &	CRR=99.58\% /FAR= 5.48 /0.41/0.49/0.41\% \\
 \cite{wilaiprasitporn2019affective}&40	&32/5&	Affective recognition&&	Raw	&CNN+LSTM/CNN+GRU \\% &	CRR= 99.9 / 99.17\%\\
 \cite{wang2019convolutional}&109/59 &	64/46 &	EO/EC/PHY/MI &&	PLV/COR	& GCNN	\\% & CRR=99.97 \%\\
%Ours   & 54 &	62 &	ERP/SSVEP/MI &&	Raw \& Connectivity metrics	& CNN+GCN	\\
\bottomrule 
\end{tabular}
\begin{tablenotes}
     \scriptsize \item  \textbf{EO}: eyes open, \textbf{EC}: eyes closed, \textbf{ERP}: event-related potential, \textbf{MI}: motor imagery, \textbf{PHY}: physical movement, \textbf{SSVEP}:steady-state visually evoked potential
\end{tablenotes}
\label{table_related_work}
\end{threeparttable}
\end{table*}
%=========================================================

\subsection{From Traditional Machine Learning to Deep Learning}

Early studies on EEG-based brain biometric for personal identification extract features such as the PSD, AR, Discrete Fourier Transform (DFT) and Wavelet Packet Transform (WPT) from EEG signals and used similarity-based classifiers to determine the identity of an individual. The similarity is measured using either Euclidean distance, Mahalanobis distance, or cross-correlation. Subsequently, k-NN and its variants, e.g., Local Large Margin Nearest Neighbors (LLMNN) are used as classifiers \cite{cai2015local, la2014human, ruiz2016cerebre}. However, those features may be varied within the same subject or similar across subjects, resulting in a decrease in performance. As a result, more complex methods known together as supervised machine learning (ML) have been developed, including Support Vector Machine (SVM), Random forest (RF), Linear Discriminant Analysis (LDA), Gaussian Mixture Model (GMM), and Multi-Layer Perceptrons (MLP) \cite{kumar2017bio,chu2017individual,bashar2016human,koike2016high}. These methods are trained on a set of data sampled to capture essential features unique to a group of individuals. Combinations of features are used to predict subjects more accurately. These methods have some limitations in that the feature extraction step requires domain knowledge and cannot be guaranteed to create entirely distinctive features.

The deep learning approaches attempt to resolve the aforementioned issues. El-Fiqi \textit{et al.} utilize the CNN to automatically extract essential features from raw EEG signals  \cite{el2018convolution}. They use the signals from Steady-State Visually Evoked Potential (SSVEP) experiments and achieve an averaged accuracy of 96.8±0.01\% for 10 subjects. In another study the CNN is concatenated with a Long-Short Term Memory (LSTM) to learn additional time-dependence information \cite{sun2019eeg}. Using a motor imagery (MI) dataset, their results have an extremely high average accuracy of 99.58\% for identifying 109 individuals.  Wilaiprasitporn \textit{et al.} \cite{wilaiprasitporn2019affective} recently expand CNN learning capability to spatial and temporal domains by performing the convoluted operation on the spatio-temporal EEG data matrices. The CNN features are then fed into Recurrent Neuron Networks (RNN), comparing LSTM and Gated Recurrent Units (GRU). According to their findings, CNN+GRU and CNN+LSTM can accurately identify 40 participants in a variety of affective states with up to 99.90\% accuracy. It indicates that both temporal and spatial information of EEG are important for this task.

\subsection{Graph Convolutional Neural Networks}

Defferrard \textit{et al.}\cite{defferrard2016convolutional}  present the convolution method for graph-domain inputs using graph theory. They transform the inputs into spectral domain then apply the K-order Chebyshev polynomials\cite{hammond2011wavelets} for graph convolution. Kipf \cite{Kipf:2016tc} proposes the GCNN, a simplified version of spectral graph convolutions in the form of a layer-wise propagation. It offers much faster training times and increased predicted accuracy. To extract more distinguishing EEG features, GCNN has been utilized to understand the underlying relationships between the EEG channels, as represented by the adjacency matrix in the graph structure \cite{song2018eeg,wang2019phase}.  Wang \textit{et al.} use a GCNN for personal identification by constructing the input graph with Pearson’s correlation (COR) and Phase-locking value (PLV) as the functional connectivity matrices \cite{wang2019convolutional}. The results reveal that the GCNN outperforms both traditional classifiers (SVM, RF, MLP) with selected input features and CNN with raw inputs. In addition, the network efficiency remains stable across brain stages, e.g., MI, eyes closed, and eyes open.

\section{Methods}
\label{sec:method}
In this section, we first describe the dataset and data pre-processing. After that, we explain graph representation of the EEG data, the proposed network architecture, and the implementation of our framework. 
%Finally, descriptions of a thorough study utilizing the proposed method is presented.

\subsection{Dataset and Pre-processing}
\label{ssec:method_data}
We conduct our experiments using the OpenBMI EEG dataset from Korea University \cite{lee2019eeg} as our benchmark. It is one of the largest dataset in terms of numbers of BCI tasks and subjects. Fifty-four healthy subjects (ages 24-35; 25 females) participated in the experiment, performing three EEG BCI tasks: Motor Imagery (MI), Event-Related Potential (ERP), and SSVEP. The EEG was acquired using the BrainAmp EEG amplifier equipped with 62 electrodes at 1 kHz sampling rate. Each subject performed each task twice on two different days (session I and II), in total yielding twice the number of trials mentioned in the BCI task descriptions below. 

\subsubsection{Motor Imagery}
In each trial, subjects were instructed to imagine moving their left hand or right hand for four seconds and then rested for approximately six seconds.
Each individual performed 100 trials, with an equal number of left-hand and right-hand trials.

\subsubsection{Event-Related Potential}
An ERP speller was utilized on a monitor. Thirty-six characters (A to Z, 1 to 9 and \_) are organized into six rows and six columns. In each session, a target character was highlighted. Following a trigger, the subjects were required to fix their gaze on the target character while the speller sequentially flashed face stimulus over speller characters 60 times and track of how many times the target character has been flashed. The procedure was repeated with different target characters until they reached all 33 characters from a predefined phrase. There were a total of 1980 trials.

\begin{subsubsection}{Steady-State Visually Evoked Potential}
Four target SSVEP stimuli were designed to flicker at the  frequencies of 5.45, 6.67, 8.57, and 12 Hz in the down, right, left, and up positions on the monitor, respectively. In each trial, the target direction was indicated with a colored stimulus. The subjects were instructed to gaze at the target direction for four seconds with a six-second gap while the stimuli flickered. Each direction appeared randomly 25 times to comprise a total of 100 trials per person. 

\end{subsubsection}

Our analysis utilizes only the data from the offline phase of the dataset. The EEG signals are bandpass filtered between 3 and 40 Hz using a $5^{\mbox{th}}$ order Butterworth filter and then downsampled to 250 Hz. Each trial of MI and SSVEP tasks has the same length of 4 seconds. Hence, the number of time samples ($T$) per trial is $T = 1000$. The ERP task is 0.8 seconds long so that $T = 200$. In addition, since the number of trials from the ERP dataset is much higher than the others, we select only the first 100 trials from each session of the ERP task so that the amount of data for all tasks is comparable. 
%As a result, the sample dimensions of ERP are $2 \times 54 \times 100  \times 62  \times 200$.

\subsection{EEG-Graph Representation} \label{sssec:EEG-Graph}
An undirected weighted graph is represented by $G = \{ V, E\}$ where $V$ indicates a set of nodes and $E$ represents a set of edges or relationships between the nodes in $V$. The edges $E$ in the graph $G$ can be represented by the weighted adjacency matrix $\mathbf{A} \in \mathbb{R}^{N \times N}$ where $N$ denotes the number of nodes in $G$. An element of $\mathbf{A}$ at the $(k,l)$ location reflects the relationship between node $k$ and node $l$.

For EEG-Graph representation, the nodes in $V$ represent the scalp electrodes, so $N=62$ in our study. Each node contains the features extracted from the raw EEG data recorded by that electrode. The weight of each edge in $E$ reflects pairwise connectivity between data from two electrodes. In this work, we investigate five connectivity measures from three aspects. The Euclidean distance measures physical connectivity between sensors. Functional connectivities are expressed in terms of signal correlation and phase synchronization. These measures are used as the elements of $\mathbf{A}$ of our graphs. 

%signals correlation using Pearson's correlation coefficient (COR), or phase synchronization using phase-locking value (PLV), phase-lag index (PLI), and $\rho$ (RHO) index. More detail about each connectivity metric is described below.

\subsubsection{Euclidean distance (DIST)} 
The Euclidean distance refers to the shortest distance between two points in the Euclidean space. In this study, the points are based on the electrode placements. Hence, the Euclidean distance $d$ between two electrodes is simply the L2 norm,
\begin{equation}
d(k,l) = ||\mathbf{p}_k-\mathbf{p}_l||_2  \label{eq:euclidean}
\end{equation}
when $\mathbf{p}_i$ and $\mathbf{p}_j$ denotes the position vectors of the $k^{\mbox{th}}$ and $l^{\mbox{th}}$ electrodes, respectively.   

\subsubsection{Pearson’s correlation coefficient (COR)}
Let $x_k(t)$ and $x_l(t),\;\; t = 1,\ldots, T$  represent the $T \times 1$ time series of EEG signals from electrodes $k$ and $l$, respectively. 
The Pearson's correlation coefficient $r \in [-1,1]$ is given by 
\begin{equation}
r(k,l) =\frac{\sum_{t=1}^{T}(x_{k}(t)-\bar{x}_{k})(x_{l}(t)-\bar{x}_{l})}{\sqrt{\sum_{t=1}^{T}(x_{k}(t)-\bar{x}_{k})^{2}}\sqrt{\sum_{t=1}^{T}(x_{l}(t)-\bar{x}_{l})^{2}}} \label{eq:pearson}
\end{equation}
where $\bar{x} = \frac{1}{T}\sum_{t=1}^T x(t)$ is the time-average of $x(t)$. The sign and magnitude of $r$ measure the direction and degree of linear association between two signals with zero time lag, respectively.  
In extreme cases, $r = -1$ reveals a complete linear inverse relationship whereas $r=1$ indicate a perfect linear relationship. A complete absence of linear dependency manifests in $r = 0$.

%\subsubsection{Phase synchronization}
Since EEG signals are made up of rhythmic oscillations, we can extract more information from phase synchronization in addition to the time-domain connectivity. The phase synchronization $\Delta\phi(k,l)$ is defined as the relative phase between the two signals $x_k(t)$ and $x_l(t)$, and is given by 
\begin{equation}
     \Delta\phi_{(k,l)} =\left | \phi_{x_{k}}(t)-\phi_{x_{l}}(t) \right| mod \text{ } 2\pi,
\end{equation}
where $mod$ denotes the modulo operation. The phase $\phi _{x}(t)$ is obtained by performing a Hilbert transform on $x(t)$.
% The relative phase of two signals is used to determine phase synchronization, $ \Delta\phi _{r}(t) =\left | \phi _{x_{i}}(t)-\phi _{x_{j}}(t) \right| mod \text{ } 2\pi $, where $\phi _{x}(t)$ is obtained by performing a Hilbert transform on the signal $x(t)$.
In this work, we extract three commonly used measures from phase synchronization \cite{wang2020brainprint}.

\subsubsection{Phase-Locked Value (PLV)}
The PLV estimates how the relative phase is distributed over the unit \cite{lachaux1999measuring}. It is expressed as
\begin{equation}
PLV(k,l) = \left | \left \langle e^{\mathrm{j}\Delta\phi _{(k,l)}(t)} \right \rangle \right |=\left | \frac{1}{T}\sum_{t=1}^{T}e^{\mathrm{j}\Delta\phi _{(k,l)}(t)} \right | \label{eq:PLV}
\end{equation}

\subsubsection{Phase-Lag Index (PLI)} This measure determines the time-lagged interdependence of two signals based on their relative \cite{stam2007phase}. It also represents the relative phase distribution; however, it is more robust to the presence of volume conduction of EEG because it neglects the distributions that center around $0$ $mod$ $\pi$. Define $sign$ as the signum function, the $PLI$ is computed from

\begin{equation}
%PLI(k,l) = \left | \left \langle sign(\Delta \phi_{(k,l)}(t)) \right \rangle \right | = \left | \frac{1}{T}\sum_{t=1}^T sign(\Delta  \phi_{(k,l)}(t) ) \right |
PLI(k,l)  = \left | \frac{1}{T}\sum_{t=1}^T sign(\Delta  \phi_{(k,l)}(t) ) \right |
\label{eq:PLI}
\end{equation}

\subsubsection{Rho index (RHO)} This index is based on Shannon\cite{rosenblum2000detection}.  It quantifies the deviation of the cyclic relative phase distribution from the uniform distribution. The discrete version is defined as 
\begin{equation}
%RHO(x_{i},x_{j}) = \frac{S_{max}-S}{S_{max}}
RHO(k,l) = 1- \frac{S}{S_{max}} \label{eq:RHO}
\end{equation}
where $S_{max} = \ln T$ is the maximal entropy of the uniform distribution quantized into $T$ bins and $S$ is the entropy of relative phase $ \Delta\phi _{(k,l)}(t)$:
\begin{equation}
S= -\sum_{m=1}^{T}p_{m}\ln(p_{m})
\end{equation}
where, $p_{m}, \;\; m = 1,\ldots, T$ is the probability of finding $ \Delta\phi _{(k,l)}(t)$  in the $m$-th bin. Its value ranges from 0 (no synchronization) to 1 (perfect synchronization).

\subsection{Graph Convolution Neural Networks}
Graph convolution neural network is commonly used due to its capabilities in learning through connections between nodes in a graph \cite{Kipf:2016tc}. It trains the network using graphs as input. The convolution operation of GCNN produces a normalized sum of features from neighboring nodes. For a graph $G$ consisting of $N$ nodes with $F$ features per node, its propagation layer is defined as

\begin{equation}
    \mathbf{H}_{n+1} = \sigma(\hat{\mathbf{D}}^{-\frac{1}{2}} \hat{\mathbf{A}} \hat{\mathbf{D}}^{-\frac{1}{2}} \mathbf{H}_n \mathbf{W}_n) \label{eq:gcnprop}
\end{equation}

where $\hat{\mathbf{A}} = \mathbf{A} + \mathbf{I}_N$; $\mathbf{A}$ is the adjacency matrix and $\mathbf{I}_N$ is the $N \times N$ identity matrix. $\mathbf{H}_n$ is an input matrix of layer $n$ and $\mathbf{H}_0$ is the initial input. $\mathbf{W}_n$ is the trainable network parameters of layer $n$. $\hat{\mathbf{D}}$ is a diagonal degree matrix of nodes used to normalize $\hat{\mathbf{A}}$. $\sigma$ is the nonlinear activation function.

\subsection{EEG-BBNet}

\begin{figure*}
\centering
  \includegraphics[width=1.0\textwidth]{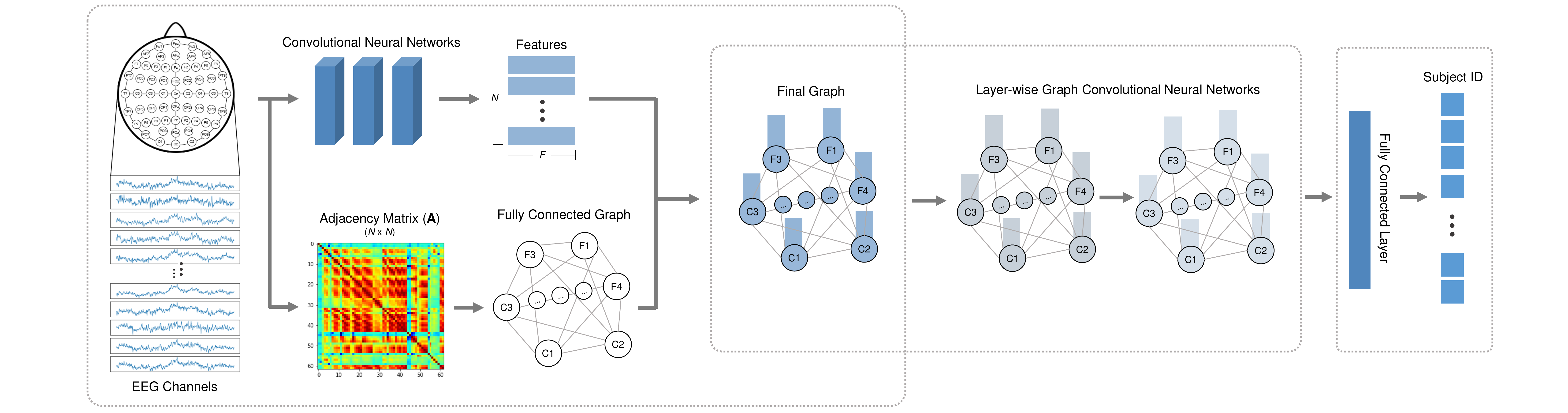}
  \caption{The Overview of EEG-BBNet.}
  \label{fig:overview}
\end{figure*}

The structure of EEG-BBNet is illustrated in \autoref{fig:overview}. It is a sequence of three components: feature extraction, graph convolution, and classification. The feature extraction employs two-dimensional depth-wise convolution (DepthwiseConv2D) layers to extract $F$ features from each electrode. The pre-processed EEG signals are fed into two DepthwiseConv2D layers with batch normalization and max pooling layers, resulting in the feature matrix of size $N \times F$. Concurrently, these EEG signals are used to compute one of the connectivity measures described in \ref{sssec:EEG-Graph} and construct the $N \times N$ weighted adjacency matrix $\mathbf{A}$. Subsequently, the feature matrix and the adjacency matrix are combined to produce the final graph that acts as the initial input for graph convolution (GC), i.e., $\mathbf{H}_0$ in (\autoref{eq:gcnprop}). This second stage contains two graph convolution (GConv) layers, whose output continues into the last segment containing the fully connected layers for personal identification. 

The detail of each layer in EEG-BBNet is summarized in \autoref{table:Configuration}. We use the Rectified Linear Unit (ReLU) as the activation function  $\sigma(\cdot)$ for every layer except the last one, which uses Softmax to identify the owner of the EEG signals. The network is optimized using $Adam$ optimizer with learning rate of 0.001 and a cross-entropy loss function as shown in \autoref{eq:entropy_loss}.

\begin{equation}
 loss =-\sum_{m=1}^{M} y_{m}\log \hat y_{m}   
\label{eq:entropy_loss}
\end{equation}

where $M$ is the number of classes or people among whom we need to identify. $y_m$ and $\hat{y}_m$ are the ground truth and the probability of classifying the data to the $m^{th}$ class, respectively. The predicted subject ID is selected to be one with the highest probability $\hat{y}_m$.  The training iteration is terminated if the validation loss does not decrease for 20 consecutive epochs. Keras framework (TensorFlow backend) is used to implement the proposed structure. We set data batch size to 32 trials. The training process is performed using NVIDIA Tesla v100 GPU with 32GB memory.

%%%%%%%%%%%%%%%%%%%%%%%%%%%%%%%%%%%%%%%%%%%%%%%%%%%%%%%%%%%%%%%
\begin{table}
\footnotesize
\centering

\caption{EEG-BBNet Model Configuration.}
\label{table:Configuration}
\begin{tabular}{lcc}
\toprule
 \textbf{Block} & \textbf{Layer} & \textbf{Configuration}\\
\midrule 
Feature Extraction  & DepthwiseConv2D  & 1 filter, kernel size 64\\
 & BatchNormalization  & default\\
  & MaxPooling2D  & pool size 32\\
& DepthwiseConv2D & 1 filter, kernel size 64\\
 & BatchNormalization  & default\\
   & MaxPooling2D  & pool size 32\\
  & Dropout  & rate 0.2\\
\midrule 
Graph Convolution  & GConv & kernel size 64  \\
  & Dropout  & rate 0.2\\
& GConv & kernel size 32 \\
  & Dropout  & rate 0.2\\

\midrule 
Classification & Flatten & \\
& Dense &  256 \\
& Dropout &  rate 0.2\\
& Dense &  128 \\
& Softmax &  $M$ \\
\bottomrule 
\end{tabular}
\end{table}
%%%%%%%%%%%%%%%%%%%%%%%%%%%%%%%%%%%%%%%%%%%%%%%%%%%%%%%%%%%%%%%

\section{Results}
\label{sec:result}
%Experimental results are reported separately for each experiment. Then all of them are summarized at the end of the section.

\begin{figure*}
\centering
  \includegraphics[width=0.8\textwidth]{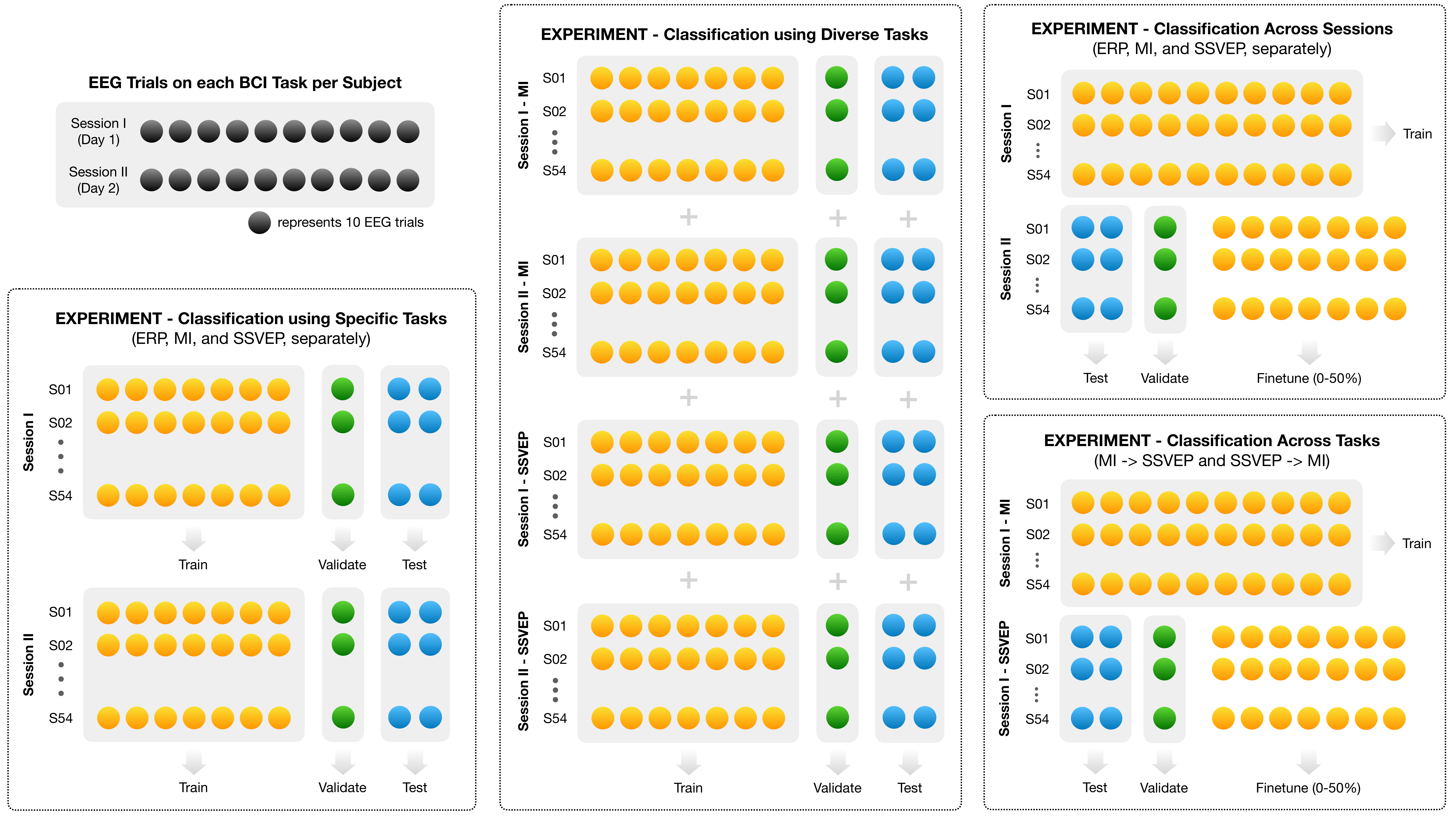}
  \caption{Splitting  training/testing/validation set for each experiment with stratified k-fold cross-validation for the classification models.}
  \label{fig:traintestsplit}
\end{figure*}

To assess the performance of our proposed method, we employ two sessions of data from each of the three BCI tasks (MI, SSVEP, ERP). The details of each experiment are described in the subsections herein and demonstrated in \autoref{fig:traintestsplit}. In all experiments, we run five-fold cross-validation with the 70:10:20 ratio for training, validation, and test datasets, respectively. We compute the Correct Recognition Rate (CRR) using \autoref{eq:CRR}, 
% the ratio of the number of identities correctly classified to the total number tested, 
and report its means and standard deviations for each case.
\begin{equation}
CRR = \frac{\text{Number of correct predictions }}{\text{Total number of predictions}} \label{eq:CRR}
\end{equation}

\subsection{EEG-Graph Representation}
\begin{figure}
\centering
\begin{subfigure}[b]{0.35\linewidth}
    \centering
  \includegraphics[width=0.95\textwidth]{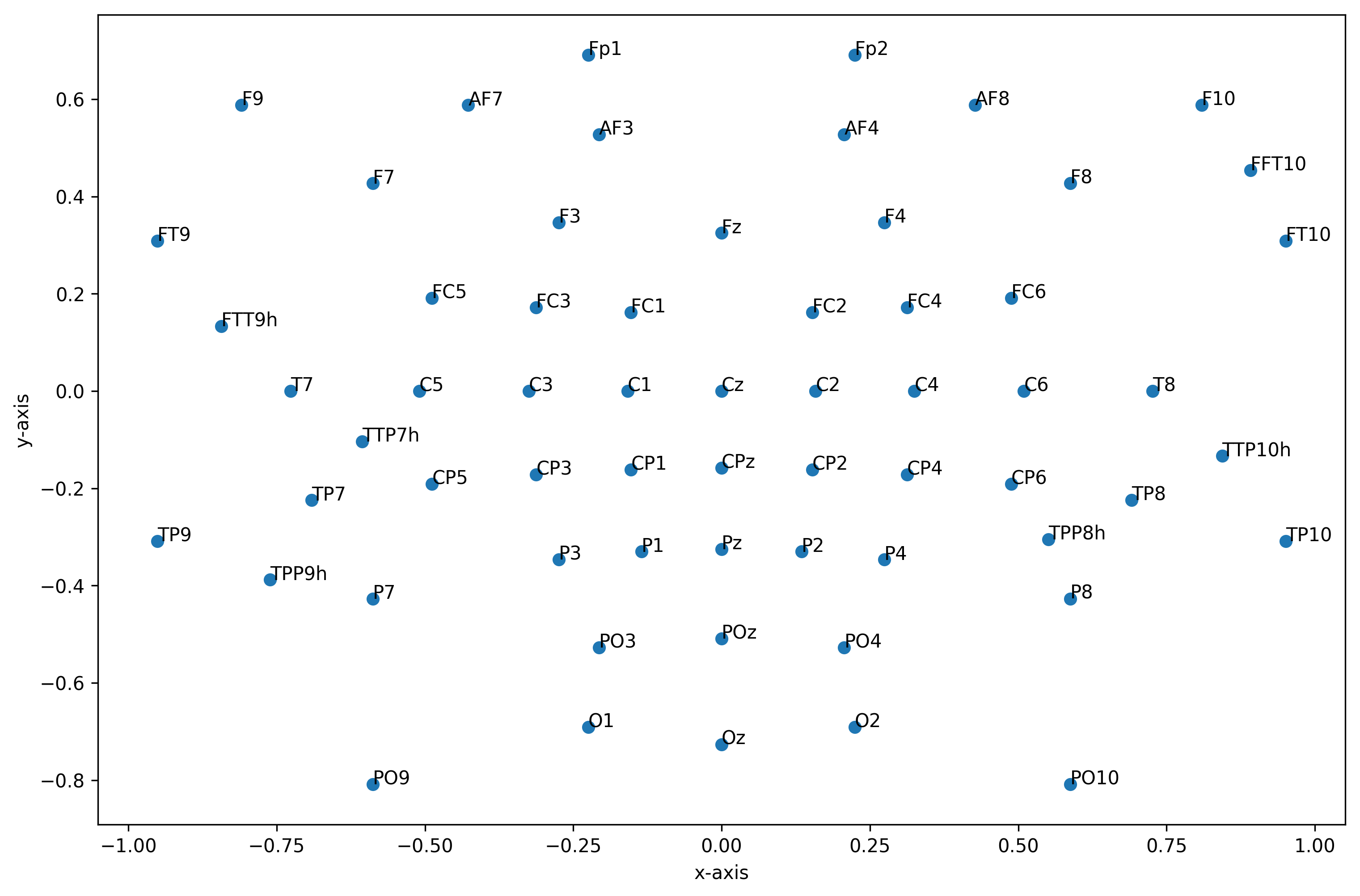}
  \caption{}
  \label{fig:a}
\end{subfigure}
\begin{subfigure}[b]{0.35\linewidth}
  \centering
  \includegraphics[width=0.95\textwidth]{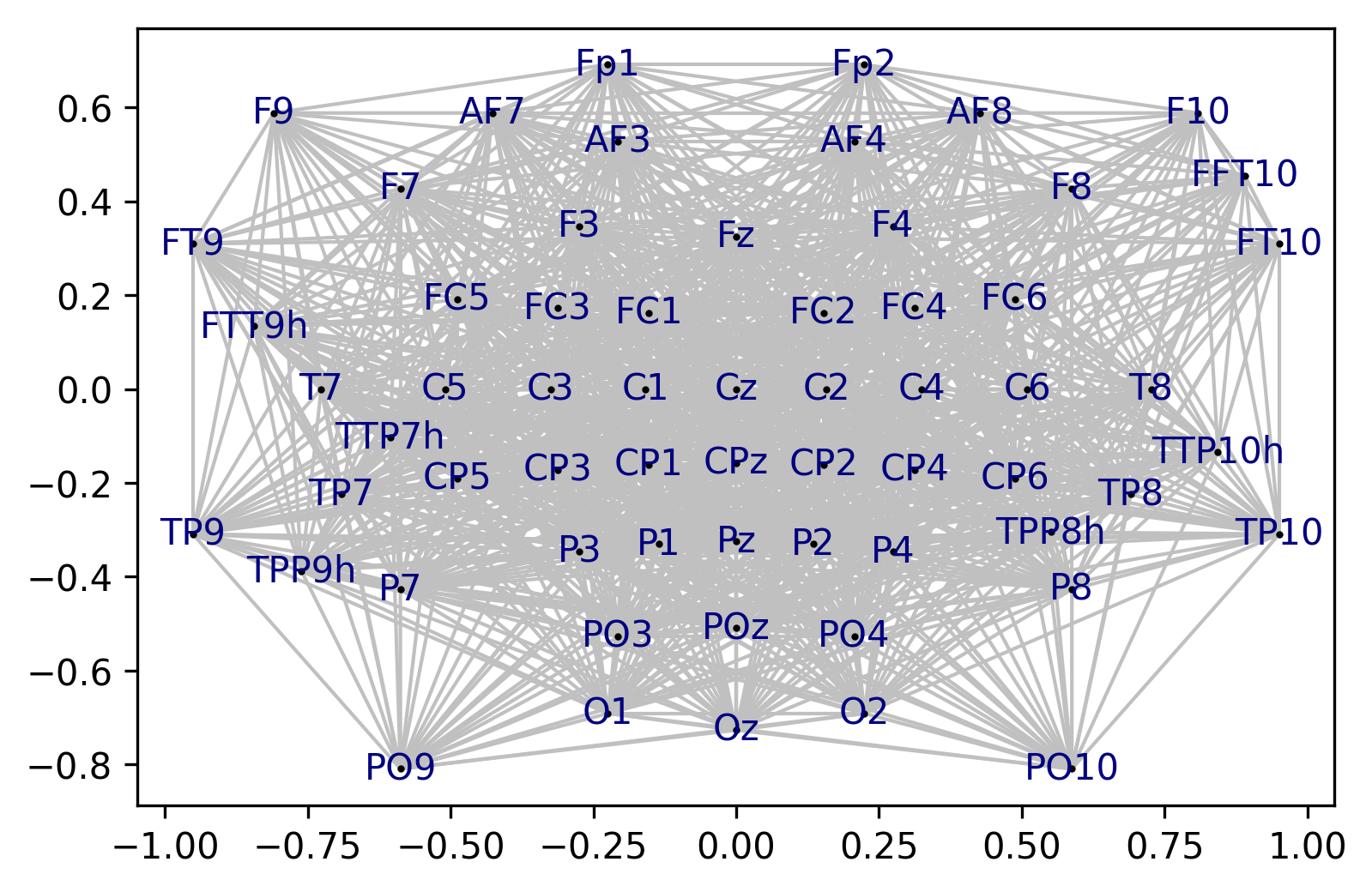}
  \caption{}
  \label{fig:b}
  \end{subfigure}
\begin{subfigure}[b]{0.25\linewidth}
  \centering
  \includegraphics[width=0.85\textwidth]{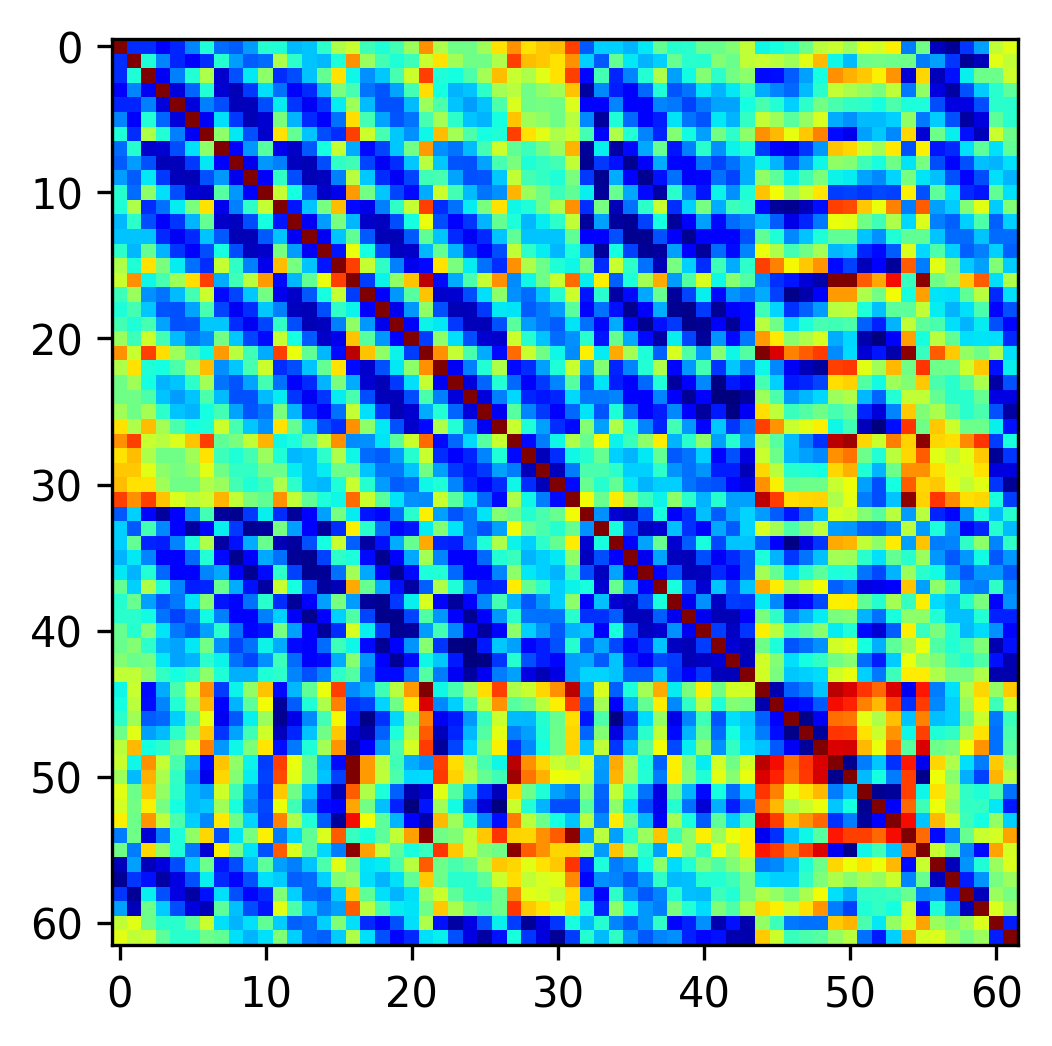}
   \caption{}
  \label{fig:c}
\end{subfigure}

 \begin{subfigure}[b]{0.228\linewidth}
  \centering
  \includegraphics[width=1\textwidth]{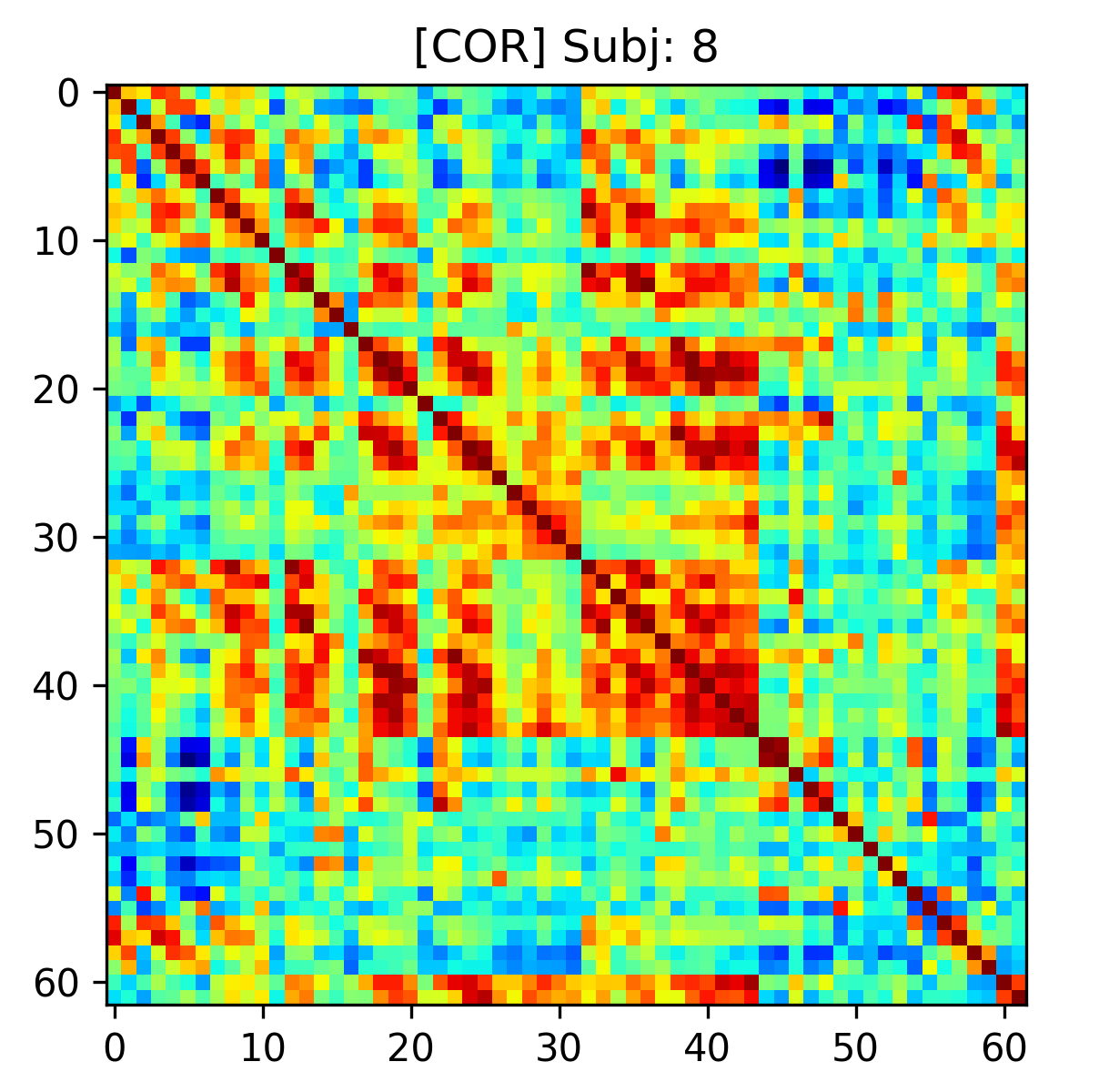}
   \caption{}
  \label{fig:d}
\end{subfigure}
\begin{subfigure}[b]{0.228\linewidth}
  \centering
  \includegraphics[width=1\textwidth]{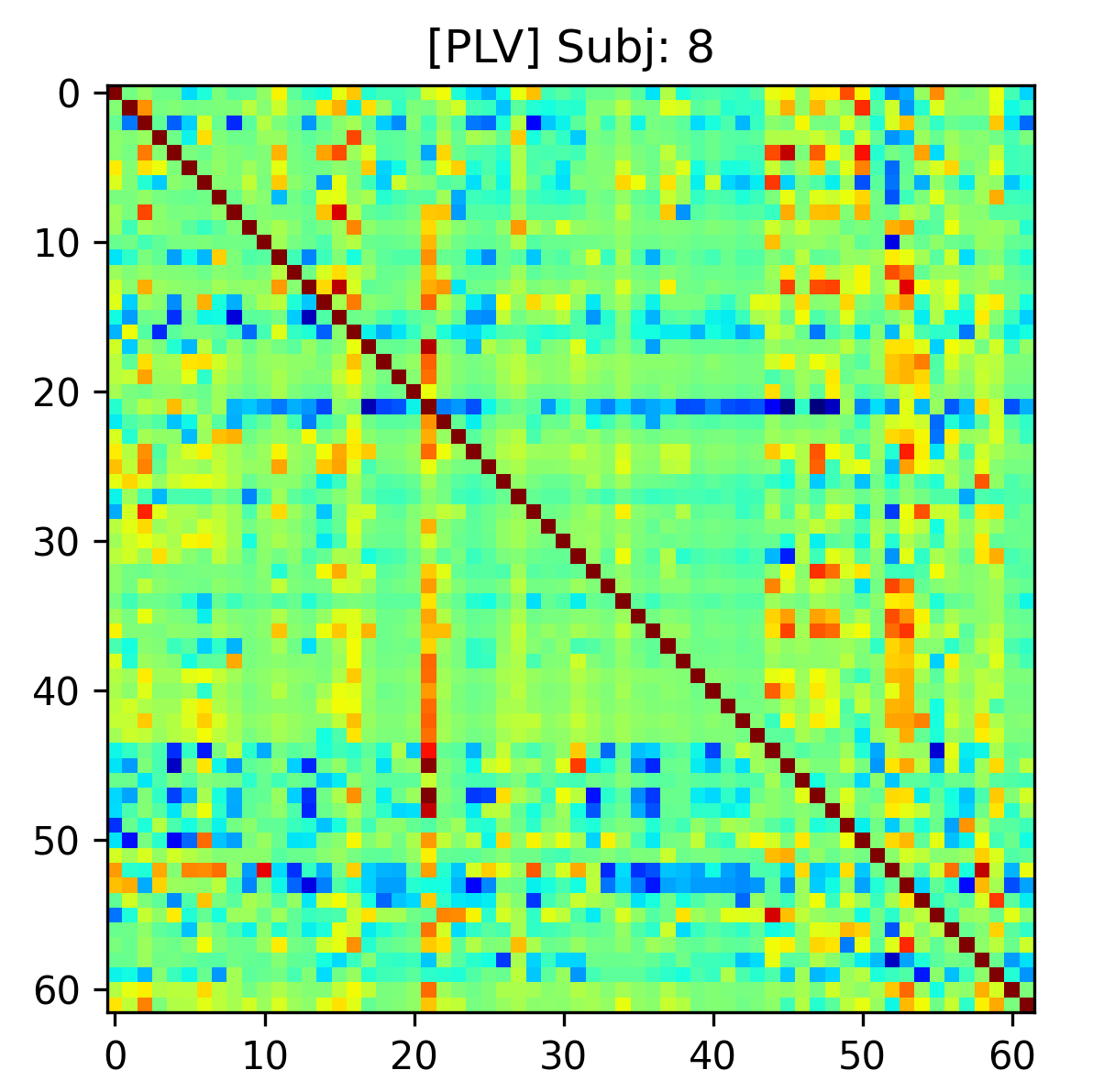}
   \caption{}
  \label{fig:e}
\end{subfigure}
\begin{subfigure}[b]{0.228\linewidth}
  \centering
  \includegraphics[width=0.95\textwidth]{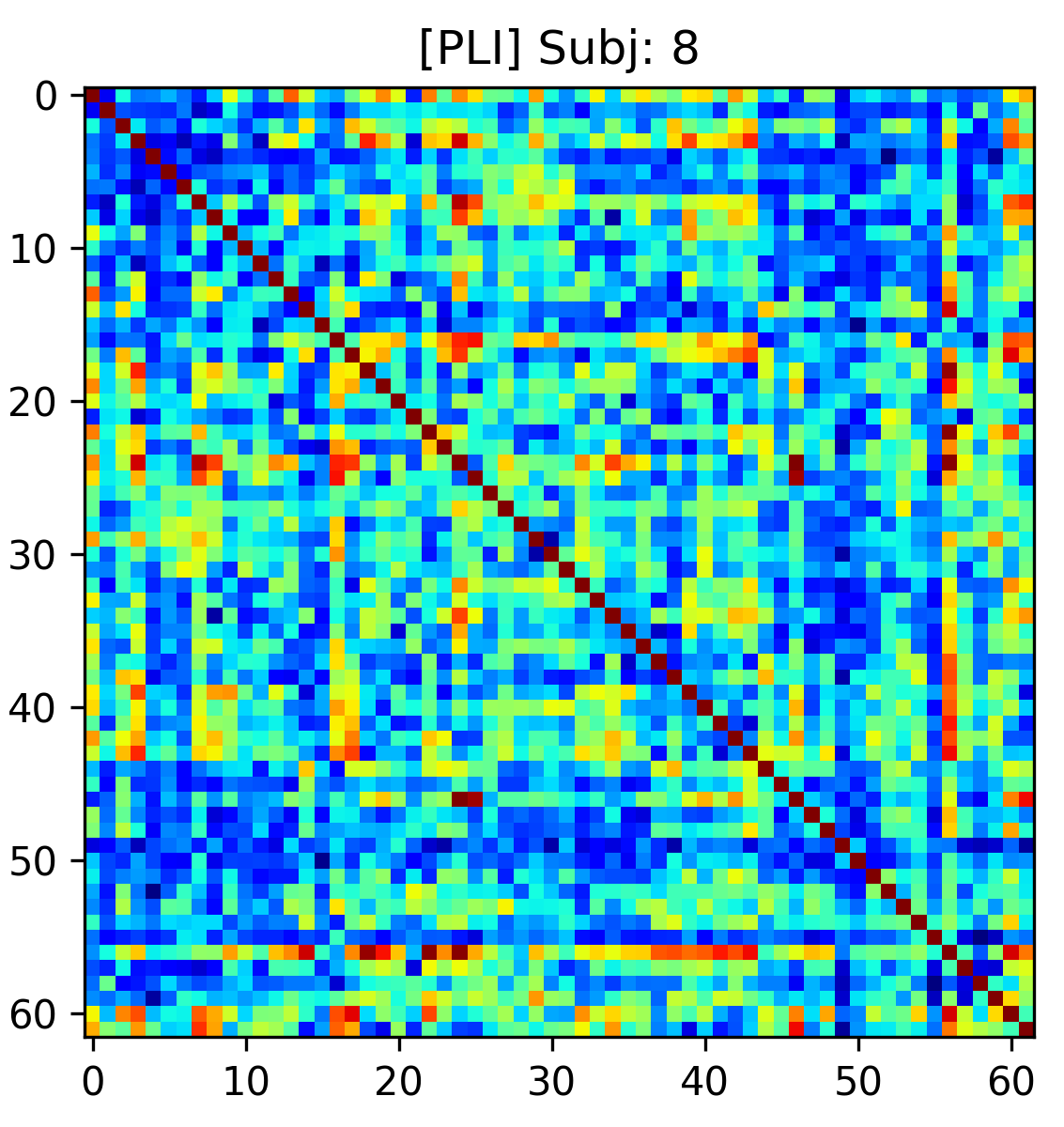}
   \caption{}
  \label{fig:f}
\end{subfigure}
\begin{subfigure}[b]{0.228\linewidth}
  \centering
  \includegraphics[width=1\textwidth]{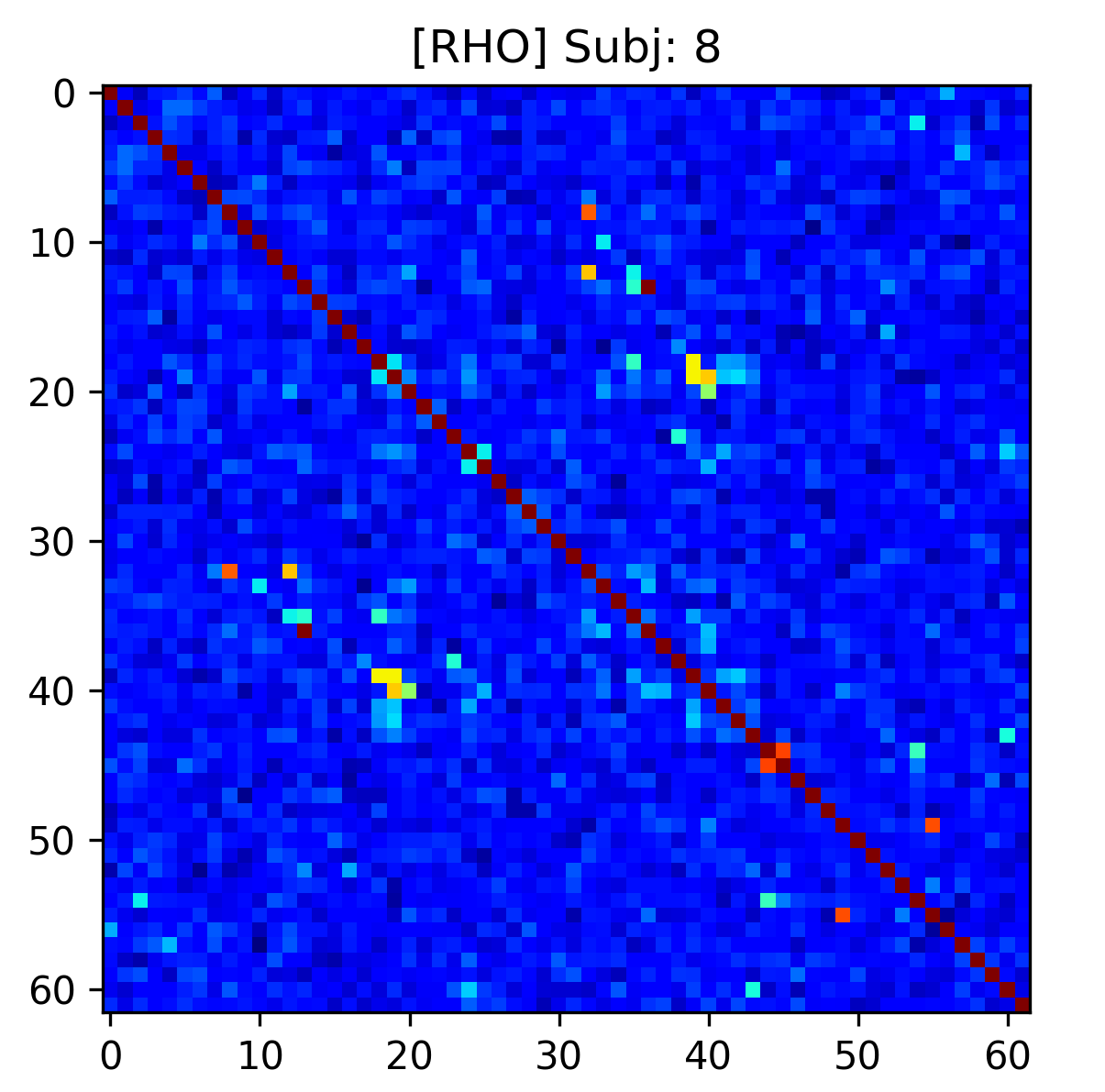}
   \caption{}
  \label{fig:g}
\end{subfigure}
  
  \caption{
        \subref{fig:a} Electrode placements projected onto 2D space,
        \subref{fig:b} Fully connected graph between EEG electrodes,
        \subref{fig:c} Euclidean distance-based adjacency matrix,
        \subref{fig:d}-\subref{fig:g} Examples of COR-, PLV-, PLI-, and RHO-based adjacency matrices of Subject 8.
    }
    \label{fig:Example_connectivity}
\end{figure}

%Sixty two channel of EEG ($Fp_{1}$, $Fp_{2}$, $F_{7}$, $F_{3}$, $F_{z}$, $F_{4}$, $F_{8}$, $FC_{5}$, $FC_{1}$, $FC_{2}$, $FC_{6}$, $T_{7}$, $C_{3}$, $C_{z}$, $C_{4}$, $T_{8}$, $TP_{9}$, $CP_{5}$, $CP_{1}$, $CP_{2}$, $CP_{6}$, $TP_{10}$, $P_{7}$, $P_{3}$, $Pz$, $P_{4}$, $P_{8}$, $PO_{9}$, $O_{1}$, $O_{z}$, $O_{2}$, $PO_{10}$, $FC_{3}$, $FC_{4}$, $C_{5}$, $C_{1}$, $C_{2}$, $C_{6}$, $CP_{3}$, $CP_{z}$, $CP_{4}$, $P_{1}$, $P_{2}$, $PO_{z}$, $FT_{9}$, $FTT_{9h}$, $TTP_{7h}$, $TP_{7}$, $TPP_{9h}$, $FT_{10}$, $FFT_{10}$, $TP_{8}$, $TPP_{8h}$, $TTP_{10h}$, $F_{9}$, $F_{10}$, $AF_{7}$, $AF_{3}$, $AF_{4}$, $AF_{8}$, $PO_{3}$, $PO_{4}$, ) illustrated in \autoref{fig:EEG_map}
As shown in \autoref{fig:Example_connectivity}(a), sixty-two EEG channels are mapped onto a normalized two-dimensional space. The channels are connected to form a graph, as illustrated in  \autoref{fig:Example_connectivity}(b). Examples of the two-dimensional adjacency matrices formed with the connectivity measures in \autoref{sssec:EEG-Graph} are then computed using a single data trial of Subject 8, as visualized in \autoref{fig:Example_connectivity}(c)-(g). The values in each matrix are normalized to [-1, 1].

Since the actual location of each electrode is not recorded during data acquisition, the electrode locations are referenced to the international 10-20 system standard. As a result, the adjacency matrices using the Euclidean distance (\autoref{fig:Example_connectivity}(c)) of every subject are identical. In contrast, the values in \autoref{fig:Example_connectivity}(d)-(g)  vary among individuals. We can observe the differences between each type of measures, representing distinct knowledge fed into the network.

To form the final graph, the adjacency matrix is combined with the feature matrix produced by the Conv2D layers of the CNN. The input size of each task differs, and so is the number of features for each task. The dimension of the feature matrix is $62 \times 12$ for ERP and $62 \times 812$ for MI and SSVEP. This graph is fed as the initial input $\mathbf{H}_0$ into the GC component of EEG-BBNet.

\subsection{Classification Accuracies of the EEG-BBNet} 

\subsubsection{Classification using specific tasks}
We train and evaluate the network for each BCI task using data from the same session, yielding two distinct findings for session I and II. For direct comparison, we implement three categories of baselines following \cite{wang2019convolutional}. Firstly, for the traditional techniques we extract the commonly used AR and PSD features and apply them to SVM and RF classifiers, respectively. Secondly, we use solely the EEG signal time series as inputs for classic DL methods (CNN and CNN+LSTM). Thirdly, for graph-based approach we construct the input graph similar to the study in \cite{wang2019convolutional}. However, we employ all five functional connectivity matrices described in \autoref{sec:method} and train the graph on the GC component of  EEG-BBNet. We refer to this model as $GCN$ for the rest of the paper.

\autoref{table_1st_result} summarizes the classification accuracy results. It reports the mean $\pm$ standard deviation of the CRR after five-fold cross validation. For the ERP task, our proposed structure identifies subjects with the highest CRRs. Among EEG-BBNet with different connectivity measures, using the Pearson's correlation coefficient achieves the best performance of 99.26 $\pm$ 0.41\% for session I while using the RHO index attains the best accuracy of 98.89 $\pm$ 0.51\% for session II. In contrast, for MI and SSVEP experiments, the best CRRs are obtained from the CNN and CNN+LSTM models. It should be noted, however, that the CNN+LSTM algorithm takes significantly longer to complete than the others since the number of model parameters is much higher.  Among the graph-based models, the one that achieves the best CRRs applies either COR or RHO as the connectivity metric. Compared to the GCN, our EEG-BBNet increases the average CRRs for every connectivity measures up to 20.8\%. 

The advantage of exploiting connectivity within the data is examined by replacing the adjacency matrix with i) an identity matrix (IDN), which represents no connectivity between electrodes, and ii) a matrix with random values (RDM). As it turns out, the model cannot learn using the RDM adjacency matrix while the IDN performs worst among the different connectivity measures. This indicates that the connectivity between channels certainly aids in the performance improvement.

\begin{table*}
\centering
\scriptsize
\begin{threeparttable}
\caption{Classification Accuracies (Mean ± SD of CRR after 5-Fold CV)(\%).} 
\begin{tabular}{l c c c  c c  c c c  c  }
\toprule
 & \multicolumn{3}{c}{\textbf{Session I}}   &\multicolumn{3}{c}{\textbf{Session II}} \\%& &\multicolumn{3}{c}{\textbf{Session I+II}} \\
 
\textbf{Method} & \textbf{ERP} &	\textbf{MI} & \textbf{SSVEP} & \textbf{ERP} & \textbf{MI} & \textbf{SSVEP} \\%& \textbf{MI} & \textbf{ERP} & \textbf{SSVEP}\\
\midrule 

PSD+AR+SVM & 91.93 $\pm$ 0.41 & 84.61 $\pm$ 0.71 & 86.15 $\pm$ 0.83 & 91.30 $\pm$ 0.73 & 84.56 $\pm$ 1.00 & 84.02 $\pm$ 0.68 \\
PSD+AR+RF & 91.85 $\pm$ 0.65 & 85.41 $\pm$ 1.45 &85.02 $\pm$ 0.85  & 89.61 $\pm$ 1.09 & 84.33 $\pm$ 1.30 & 80.91 $\pm$ 1.09 \\
Raw+CNN         & 76.26 $\pm$ 3.95  & \textbf{93.52 $\pm$ 1.20} & \textbf{93.04 $\pm$ 1.72} & 73.83 $\pm$ 5.38 & 92.41 $\pm$ 0.27 & 89.63 $\pm$ 1.21 \\
Raw+CNN+LSTM    & \textbf{98.70 $\pm$ 0.92}  & 92.83 $\pm$ 1.31 & 90.59 $\pm$ 1.67 & \textbf{98.20 $\pm$ 0.33} & \textbf{92.41 $\pm$ 1.32} & \textbf{89.65 $\pm$ 1.93} \\
\midrule 
GCN[COR]  & 93.89 $\pm$ 0.40  & \textbf{90.14 $\pm$ 0.68} & 86.48 $\pm$ 5.91 & \textbf{92.57 $\pm$ 2.18} & \textbf{90.80 $\pm$ 0.75} & \textbf{85.77 $\pm$ 1.18} \\
GCN[PLV]  & 78.09 $\pm$ 5.71 & 79.80 $\pm$ 3.07 & 73.39 $\pm$ 6.35 & 70.48 $\pm$ 8.07 & 80.19 $\pm$ 3.39 & 67.19 $\pm$ 5.60 \\

GCN[PLI]  & 90.46 $\pm$ 1.38  & 84.13 $\pm$ 0.58 & 83.24 $\pm$ 0.52 & 86.57 $\pm$ 3.16 & 88.44 $\pm$ 1.39 & 64.50 $\pm$ 3.90 \\

GCN[RHO]  & \textbf{94.11 $\pm$ 1.07}  & 86.80 $\pm$ 6.60 & \textbf{86.62 $\pm$ 4.41} & 90.37 $\pm$ 1.04 & 85.65 $\pm$ 3.51 & 81.70 $\pm$ 4.33 \\
\midrule 
EEG-BBNet[COR]      & \textbf{99.26 $\pm$ 0.41} & \textbf{91.20 $\pm$ 1.13} & \textbf{89.20 $\pm$ 2.65} & 98.17 $\pm$ 0.71 & \textbf{91.50 $\pm$ 1.24} & 86.00 $\pm$ 2.28 \\
EEG-BBNet[PLV]      & 98.89 $\pm$ 0.34 & 85.54 $\pm$ 1.52 & 83.39 $\pm$ 2.95 & 97.56 $\pm$ 0.56 & 87.59 $\pm$ 1.97 & 79.19 $\pm$ 4.37 \\
EEG-BBNet[PLI]      & 98.72 $\pm$ 0.74 & 87.72 $\pm$ 1.49 & 86.13 $\pm$ 2.20 & 97.98 $\pm$ 0.50 & 88.76 $\pm$ 2.31 & 81.44 $\pm$ 1.42 \\
EEG-BBNet[RHO]      & 99.07 $\pm$ 0.33 & 89.54 $\pm$ 2.38 & 88.28 $\pm$ 2.48 & \textbf{98.89 $\pm$ 0.51} & 88.91 $\pm$ 3.23 & \textbf{87.11 $\pm$ 3.47} \\
EEG-BBNet[DIST]     & 98.87 $\pm$ 0.30 & 91.11 $\pm$ 0.80 & 89.11 $\pm$ 1.95 & 98.59 $\pm$ 0.25 & 89.72 $\pm$ 4.69 & 87.04 $\pm$ 2.76\\
EEG-BBNet[IDN]     & 98.70 $\pm$ 0.72 & 84.50 $\pm$ 5.32 & 83.26 $\pm$ 4.80 & 97.50 $\pm$ 1.15 & 83.11 $\pm$ 5.91 & 84.20 $\pm$ 3.65\\

\bottomrule 
\end{tabular}
%\begin{tablenotes}

%     \scriptsize \item  
%     \textit{Optimal values among respective groups shown in boldface}\\
%     \textbf{IDN}: Identity matrix, \textbf{DIST}: Euclidean distance, \textbf{COR}: Pearson's correlation coefficient, \\ \textbf{PLV}: Phase-locked value, \textbf{PLI}: Phase-lag index, \textbf{RHO}: Rho index.\\
     
% \end{tablenotes}
\label{table_1st_result}
\end{threeparttable}
\end{table*}

\subsubsection{Classification using diverse tasks}
It is more practical if the network can identify subjects regardless of the task. Therefore, we conduct this experiment by combining the data from session I of two BCI tasks (MI, SSVEP) and randomly separating them into the training and test sets.

\begin{figure}
\centering
  \includegraphics[width=0.45\textwidth]{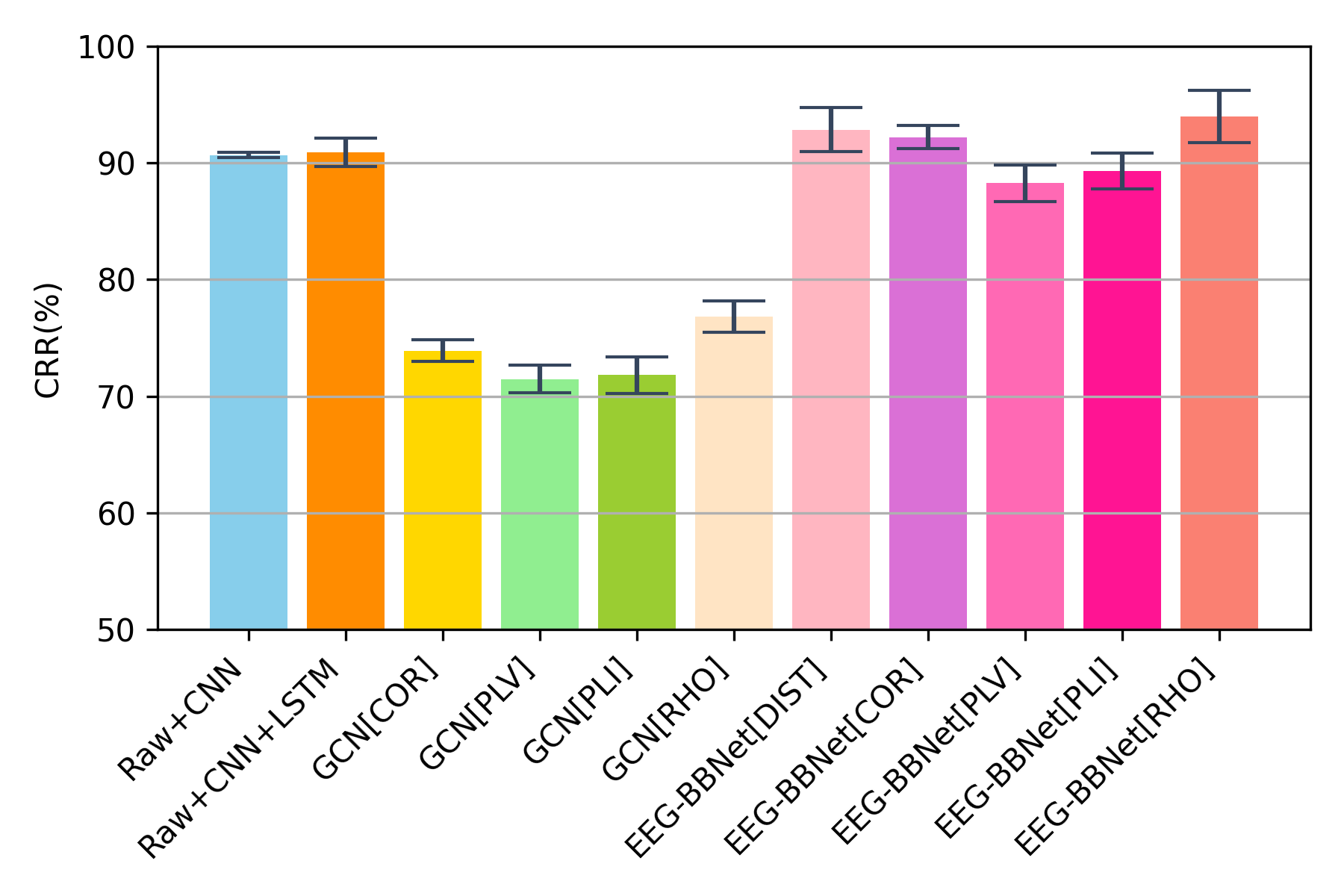}
  \caption{Classification results when using diverse tasks. (Mean ± SD Of 5-Fold CV)(\%)}
  \label{table_Sum}
\end{figure}

After we gather the MI and SSVEP data and evaluate on randomly selected samples from both tasks, EEG-BBNet with RHO index outperforms all algorithms with the CRR of 93.95 $\pm$ 2.24\%, as shown in \autoref{table_Sum}. Even though the data become more diverse with two BCI tasks, EEG-BBNet still achieves promising results. This suggests the generalizability of our network, allowing us to train on data from several tasks simultaneously.

% After we gathered all MI and SSVEP data in one place, the number and diversity of training set increase. After evaluating on test set which also included data from both tasks, as shown in \autoref{table_Sum}, our proposed EEG-BBNet with RHO outperformed all baselines with COR of 93.95 $\pm$ 2.24. Compared to Experiment III, the model can identify subjects more efficiently since it has seen data from both tasks, i.e., the test tasks is known. These refer to more generalized of our network, allowing to train on data from several tasks simultaneously.

\subsection{Practicality of EEG-BBNet}
\subsubsection{Training and testing across sessions}

In practice, a biometric system should be capable of identifying users on any days. The purpose of this experiment is to test the robustness of our system by training and evaluating the model using data from different sessions. In each BCI task, we train the classifiers using data from the first session, and test it on the data from the second session.

\begin{figure}
\centering
\begin{subfigure}[b]{1\linewidth}
    \centering
  \includegraphics[width=0.9\textwidth]{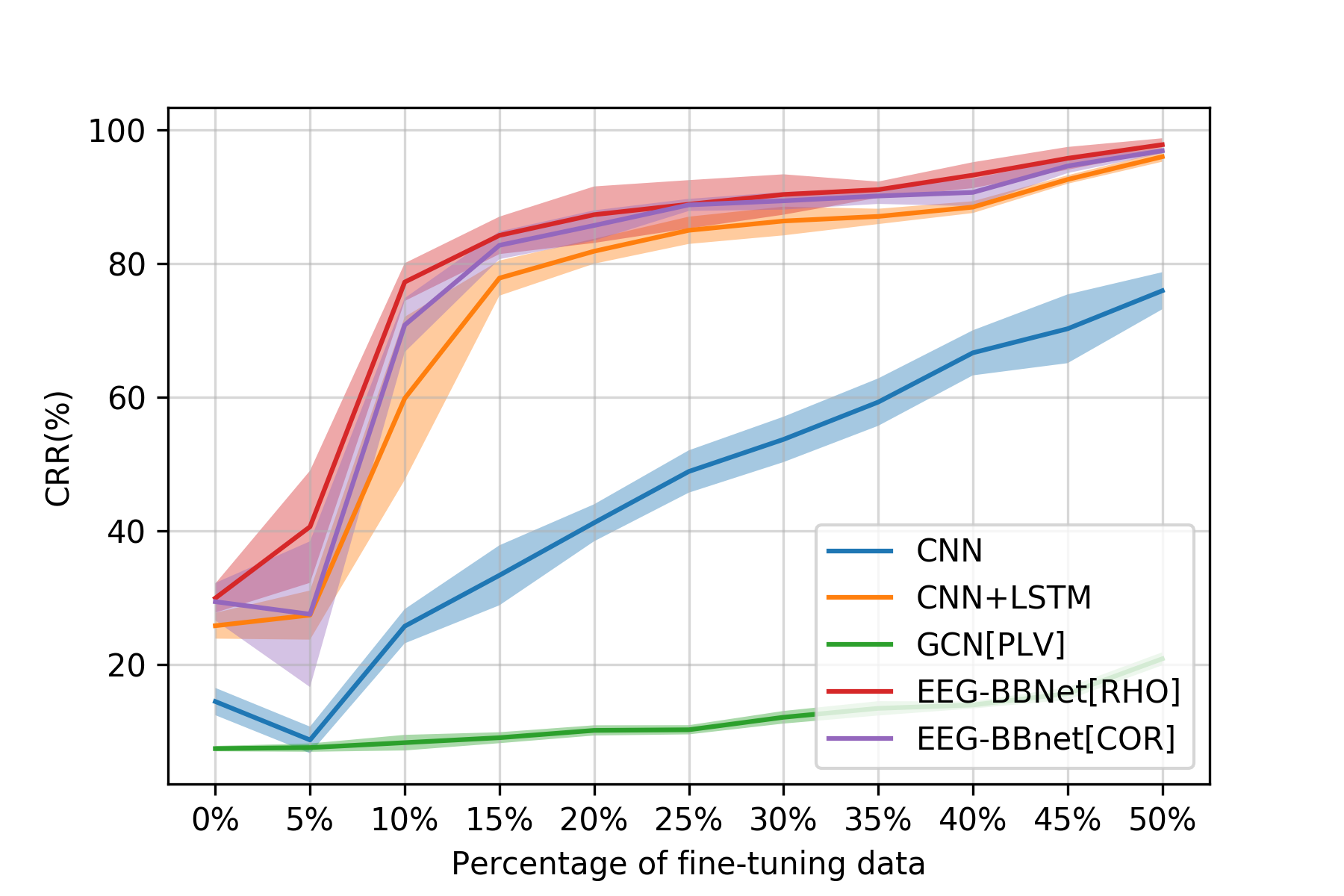}
    \caption{}
  \label{fig_ac:a}
\end{subfigure}
\begin{subfigure}[b]{1\linewidth}
  \centering
  \includegraphics[width=0.9\textwidth]{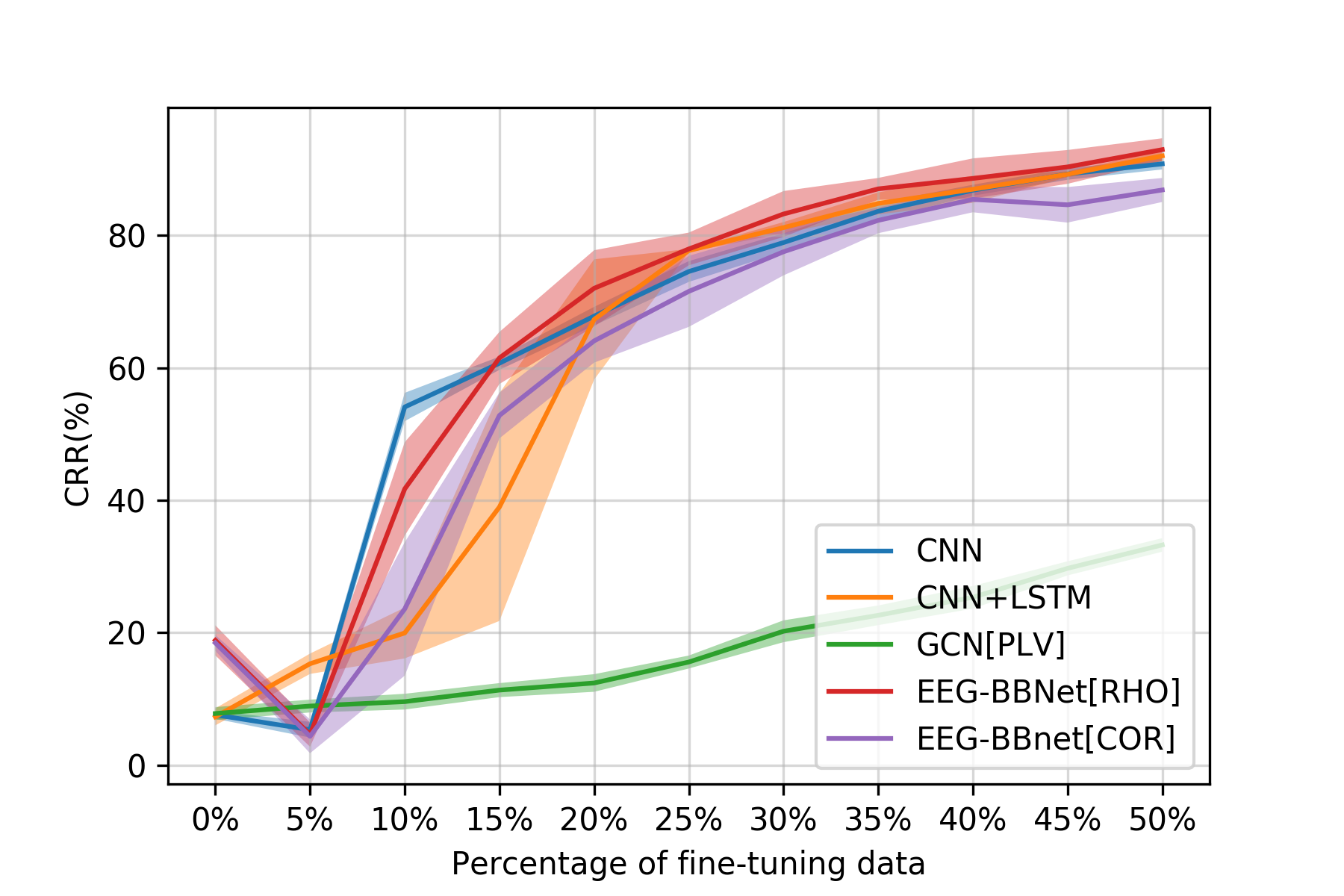}
    \caption{}
  \label{fig_ac:b}
\end{subfigure}
\begin{subfigure}[b]{1\linewidth}
  \centering
  \includegraphics[width=0.9\textwidth]{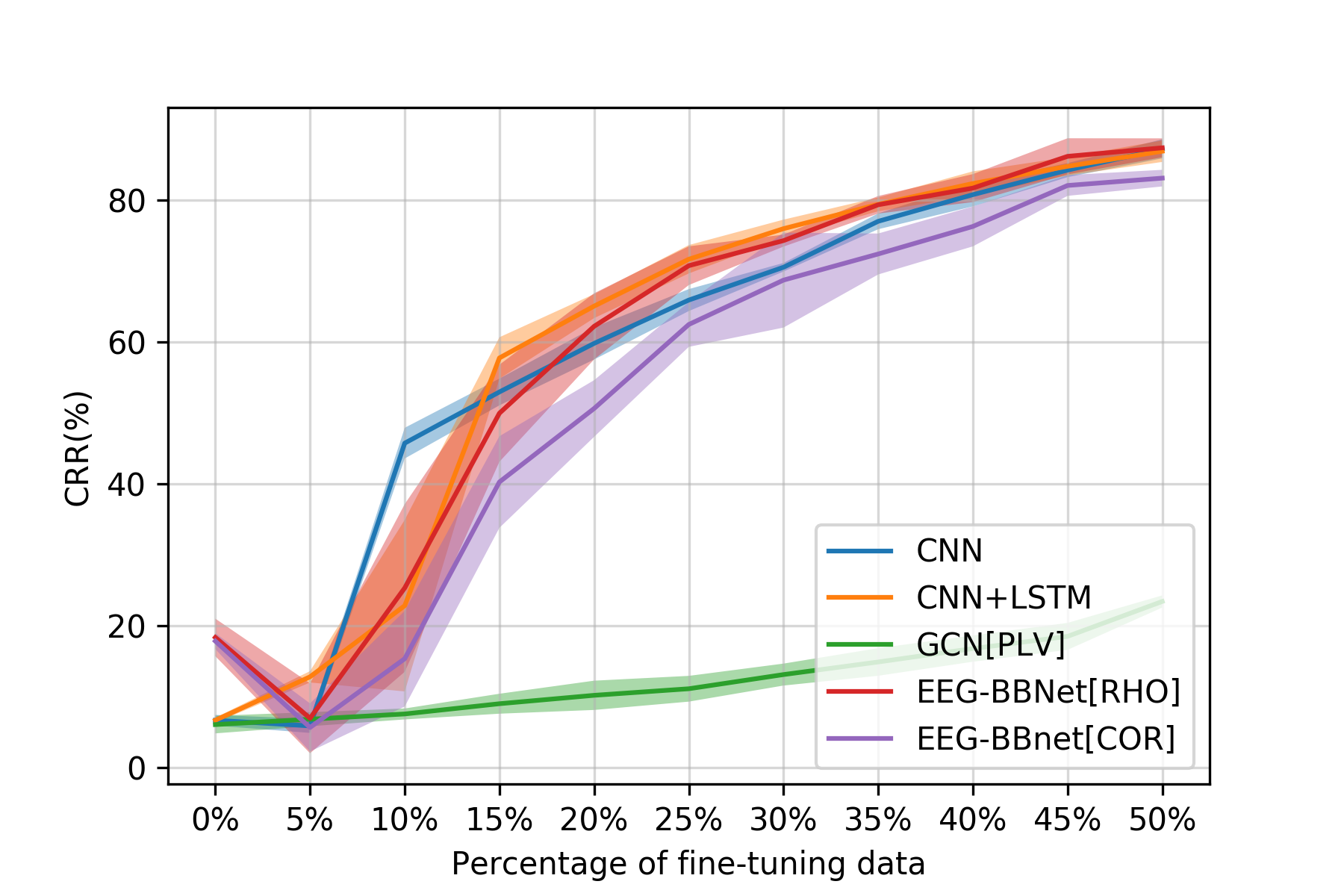}
  \caption{}
\label{fig_ac:c}
\end{subfigure}
  
  \caption{Classification results when training and testing across sessions on (\subref{fig_ac:a}) ERP, (\subref{fig_ac:b}) MI, and (\subref{fig_ac:c}) SSVEP tasks. }

  \label{fig:across}
\end{figure}

After training the model with session I data and testing it on session II data, the EEG-BBNet with RHO index obtains the best CRR of 29.98\% $\pm$ 2.17\%, 18.85\% $\pm$ 2.27\%, and 18.31\% $\pm$ 2.64\% for the ERP, MI, and SSVEP tasks, respectively. This low levels of classification accuracies are to be expected, since data from different sessions can be quite divergent. To improve the classification performance, we fine-tune the model by adding 5-50\% of the data from session II into the training set with 5\% increments. In \autoref{fig:across}, we visualize the classification results of EEG-BBNet[RHO] and EEG-BBNet[COR] against the other state-of-the-art approaches, namely CNN, CNN+LSTM, and GCN. The best connectivity measure for the GCN is PLV for this experiment.

\autoref{fig:across} illustrates that performances of all models improves with the addition of fine-tuning data. However, the improvements of EEG-BBNet and CNN+LSTM are the most evident across all BCI tasks, with the EEG-BBNet[RHO] holding a slight edge over the CNN+LSTM for the ERP task.
It is observed that the CRR of EEG-BBNet[RHO] increases substantially from 29\% to almost 80\% with only 10\% of fine-tuning data. For MI and SSVEP tasks, similar trends can be observed. In addition, the performances of EEG-BBNet[RHO], CNN+LSTM, and CNN are comparable when the amount of the fine-tuning data is greater than 35\%. Interestingly, the GCN[PLV] does not benefit from this as much as the other algorithms.

\subsubsection{Training and testing across BCI tasks}
Some BCI tasks may not be acquirable in real-world settings, and EEG signal registration for biometric purposes may be limited. As a result, data from certain tasks may be missing. Hence, it is desirable that the system is capable of identifying a person from the EEG data of untrained tasks. We investigate the efficacy of EEG-BBNet for this circumstance by training the model on MI data and evaluating on SSVEP data, and vice versa. Because MI and SSVEP have the same data dimensions, it allows us to use the same network architecture. 

\begin{figure}
\centering
\begin{subfigure}[b]{1\linewidth}
  \centering
  \includegraphics[width=0.9\textwidth]{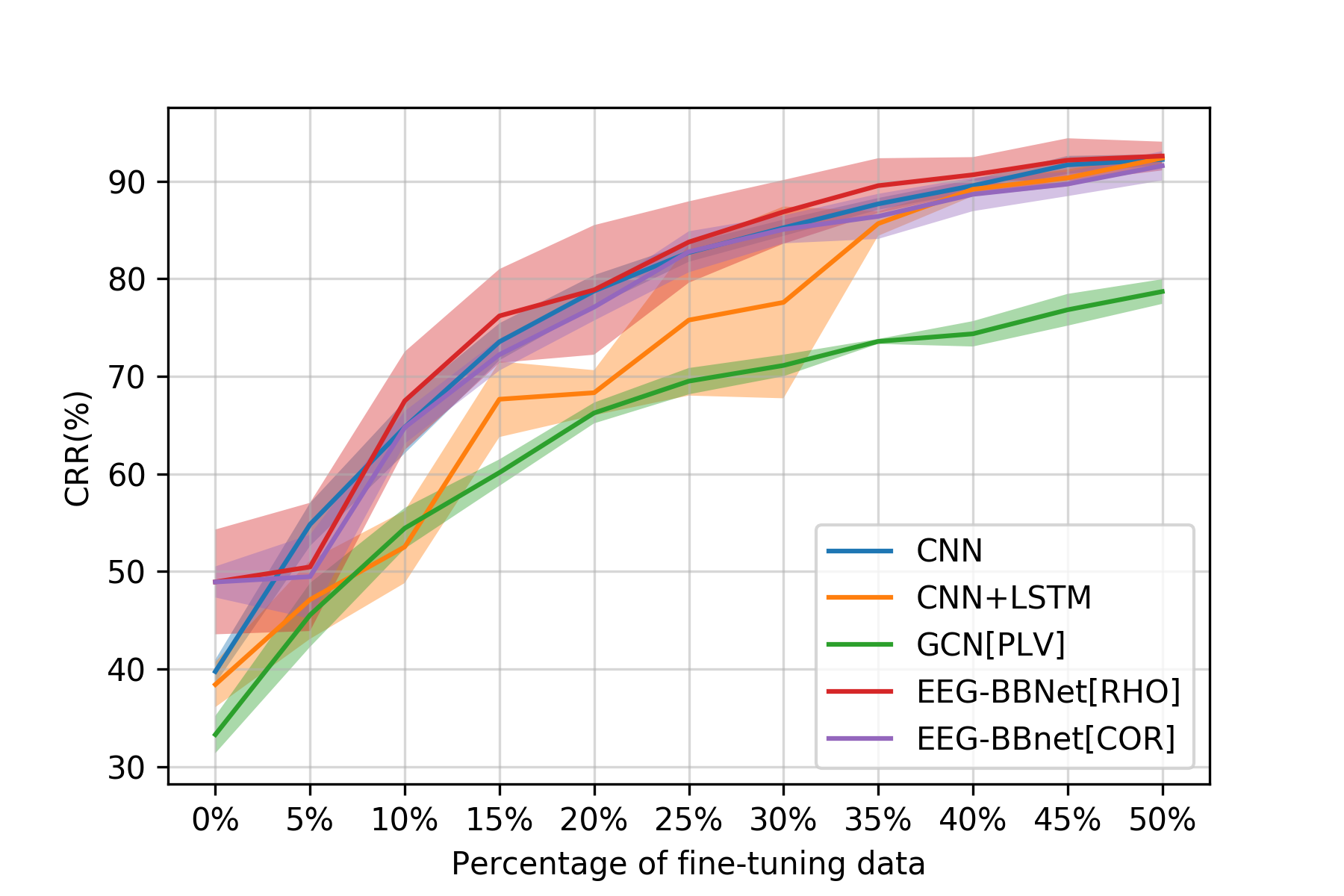}
     \caption{}
  \label{fig_acT:a}
\end{subfigure}
\begin{subfigure}[b]{1\linewidth}
  \centering
  \includegraphics[width=0.9\textwidth]{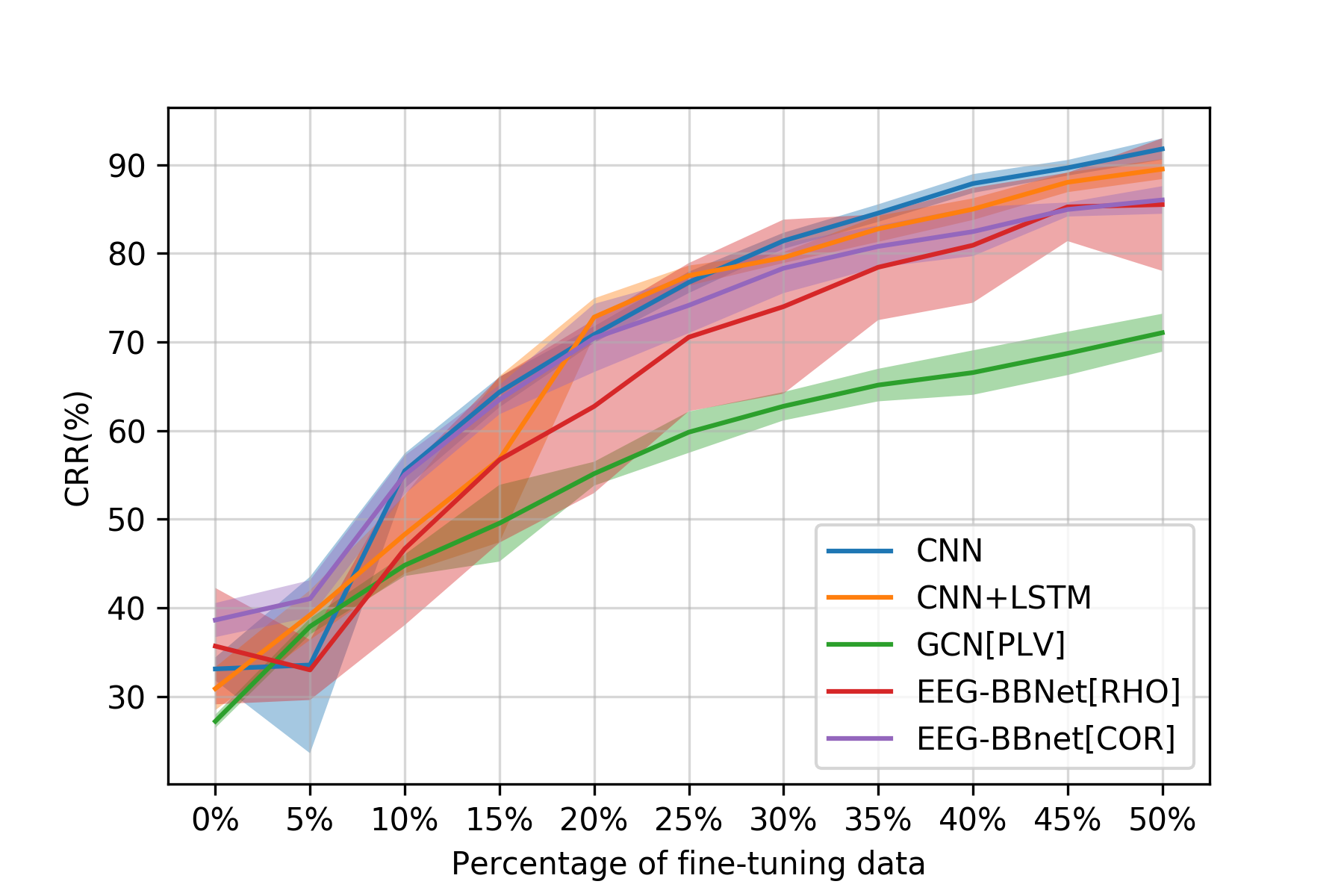}
     \caption{}
  \label{fig_acT:b}
\end{subfigure} 

  \caption{Classification results when (\subref{fig_acT:a}) training on SSVEP and testing on MI, and (\subref{fig_acT:b})training on MI and testing on SSVEP. }
  \label{fig:acrossTask}
\end{figure}

The results are depicted in \autoref{fig:acrossTask}. After training the model using SSVEP and testing on MI, EEG-BBNet with RHO index achieves the highest CRR of 48.93 \% without fine-tuning. It infers that the functional connectivity between channels in SSVEP task that is partially transferable to MI task. The classification performances of EEG-BBNet[RHO], EEG-BBNet[COR] and CNN are similar for the 5-35\% fine-tuning. With 40\% fine-tuning data, CNN+LSTM also attains similar CRR. On the other hand, when using the MI data for training and SSVEP for testing the CRRs of EEG-BBNet[COR] and CNN are similar with 5-20\% fine-tuning. When the percentage of fine-tuning data increases up to 50\%, EEG-BBNet[COR] and EEG-BBNet[RHO] perform slightly worse than CNN and CNN+LSTM.  Note that the robustness of EEG-BBNet[RHO] also decreases, as seen from its greater variance after five-fold cross-validation.  

\subsubsection{Comparison on different groups of EEG electrodes}

\begin{figure}
\centering
  \includegraphics[width=0.45\textwidth]{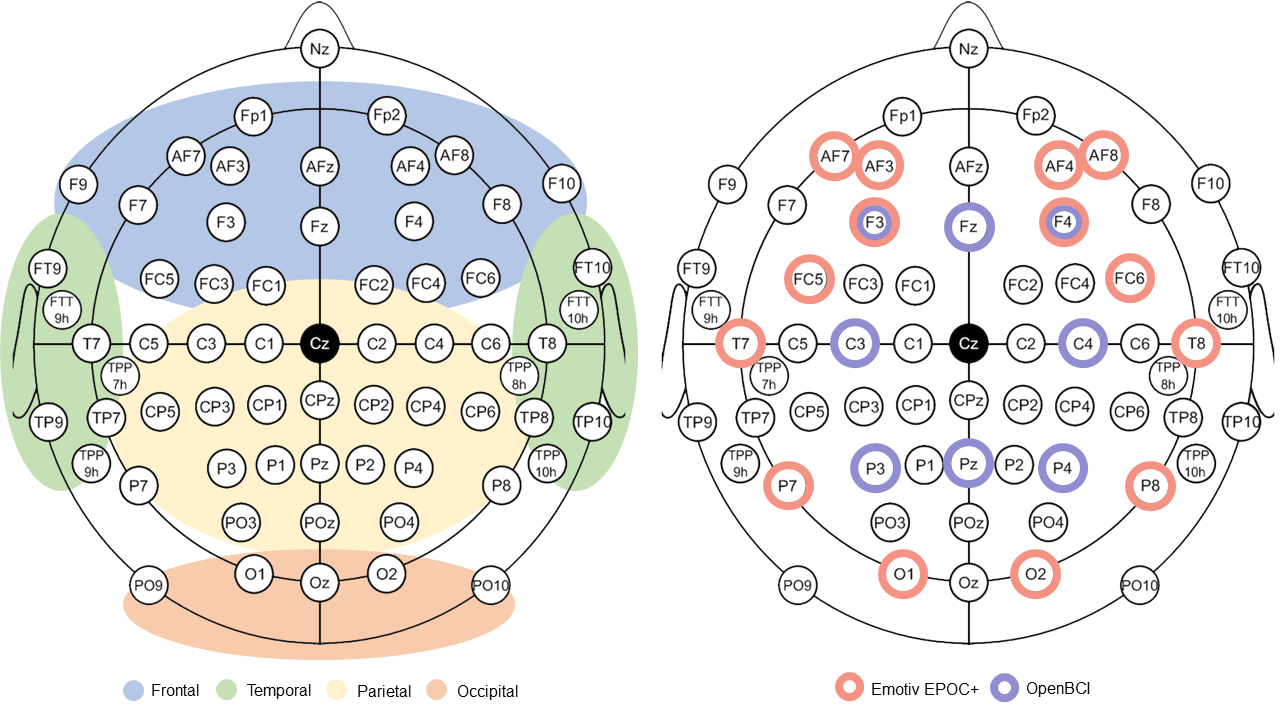}
  \caption{Different groups of EEG Electrodes clustered by brain regions (left) and consumer-grade EEG devices (right). }
  \label{fig:head}
\end{figure}

To alleviate the difficulty of EEG recordings in practice, we examine the possibility of reducing the number of EEG electrodes. Two strategies are chosen as guidelines for channel selection as illustrated in \autoref{fig:head}. The first one is based on the anatomy of the brain. The electrodes are grouped according to their placements on the four brain regions: frontal, parietal, occipital, and temporal lobes. The second selection is based on the montages of consumer-grade EEG devices, Emotiv EPOC+ and OpenBCI, which have 14 and 8 electrodes, respectively. We compare the classification results when only data from each of the aforementioned electrode subsets is used in the experiment.

In this experiment, we evaluate on the ERP data session I. \autoref{table_brainSec} presents the CRRs when we utilize different groups of EEG electrodes. The results show that the subset of data from frontal lobe electrodes achieves considerably higher CRRs for every connectivity metrics.

\begin{table*}
\centering
\scriptsize
\begin{threeparttable}
\caption{Performance Comparison when using subsets of EEG electrodes. } 
\begin{tabular}{l c c c c c}
\toprule
\textbf{Groups of EEG electrodes} & EEG-BBNet[DIST] &EEG-BBNet[COR] & EEG-BBNet[PLV] & EEG-BBNet[PLI] & EEG-BBNet[RHO]  \\
\midrule 
Frontal lobe & \textbf{95.02 $\pm$ 1.24} & \textbf{95.56 $\pm$ 1.09}  & \textbf{94.11 $\pm$ 0.95} & \textbf{93.52 $\pm$ 1.61} & \textbf{96.17$\pm$ 0.65} \\
Parietal lobe & 70.48 $\pm$ 4.00 & 72.72 $\pm$ 2.00  & 68.19 $\pm$ 2.35 & 63.33 $\pm$ 6.26  & 81.98 $\pm$ 1.21 \\
Occipital lobe &  31.46 $\pm$ 3.32 & 34.70$\pm$ 3.73  & 32.80 $\pm$ 0.97 & 28.89 $\pm$ 1.93  & 56.76 $\pm$ 3.73 \\
Temporal lobe &  68.43 $\pm$ 2.01 & 79.63 $\pm$ 1.88  & 72.41 $\pm$ 2.59 & 70.33 $\pm$ 1.80  & 71.67$\pm$ 25.45 \\

Emotiv EPOC+ (14 channels) &  73.31 $\pm$ 5.70 & 78.81 $\pm$ 3.81  & 72.50 $\pm$ 4.60 & 67.33 $\pm$ 6.52  & 82.65 $\pm$ 1.83 \\
OpenBCI (8 channels) &  72.07 $\pm$ 4.85 & 77.98 $\pm$ 1.46  & 73.35 $\pm$ 2.42 & 65.35 $\pm$ 474  & 77.65 $\pm$ 13.25 \\
BrainAmp (62 channels)&  98.87 $\pm$ 0.30 & 99.26$\pm$ 0.41 & 98.89 $\pm$ 0.34 & 98.72 $\pm$ 0.74  & 99.07 $\pm$ 0.33 \\
\bottomrule 
\end{tabular}
\label{table_brainSec}
\end{threeparttable}
\end{table*}

\section{Discussion}
\label{sec:discuss}
%\subsection{Effective of Graph Neural Networks in biometrics}
%\subsection{Significant advantages of EEG-BBNet}
Multiple studies have presented powerful tools for brain biometrics utilizing ML or DL techniques to extract features from EEG channels. Most of them have achieved encouraging results. On the other hand, the number of GCNN approaches remains limited despite their unique capability to learn relationships between graph nodes. Recent research has demonstrated the advantages of utilizing GCNN in brain biometrics. However, the features in each EEG channel are not employed. Our proposed research combine the best of both worlds by incorporating the automatic feature extraction of CNN and channel connection knowledge of GCNN, thus providing the model with substantially more meaningful information.
To accomplish this, we extract features from each channel using CNN and passe them into the GCNN as features of each graph node. This fusing of features results in higher classification performance, as demonstrated in our experiments. 

%\subsection{Analysis of graph functional connectivity}
 We use a variety of connectivity metrics to capture different aspects of relationships between data from different electrodes. 
 From the experimental results, the Pearson's correlation coefficient and RHO index performs best among the measures. Since it is well-known that data from nearby electrodes are highly dependent due to both the volume conduction of the brain and active reference electrode, COR apparently captures these correlations.  RHO index, on the contrary, extracts phase information from the oscillating behavior of the EEG signals. This suggests that phase synchronization is also an essential information for distinguishing between individuals. 

The benefit of employing the relationship between channels is validated by defining a trivial graph using the identity matrix. This graph represents that all channels are independent. The classification accuracy diminishes from when the relationships are specified. We also explore the case of random connectivity and observe that the system does not converge at all. Therefore, we are confident that meaningful functional connectivity plays a vital role for EEG-BBNet.

Regarding real-world implementation, the results confirm the difficulty associated with applying the model trained on one day directly to data from a different day. However, our model demonstrates superior adaptability by utilizing fewer samples for fine-tuning. It may be inferred that the relationship between channels in the same individual remains similar over time, necessitating a smaller sample size from an incoming session to learn from. Having only one training and one testing session/task limits our ability to measure adaptability. Nevertheless, we believe our model would be more adaptable to unseen data if trained with data from multiple sessions. Adapting to an unknown task, however, is highly challenging and requires further studies.

%From C1 experiment, the results suggested that the CNN and CNN+LSTM models might retain enormous information on the EEG patterns of each individual, resulting in the model performing poorly when tested on data from an unseen session. On the other hand, additional information regarding the relationship between channels makes the EEG-BBNet more adaptable to the new session and achieved better results than others.

%\subsection{Applicability and Adaptability}
Any biometric system is better received when it is compact 
with minimal hardware requirement. Therefore, we examined the performance of the system when fewer electrodes are used. The selection of electrodes is based on either the functional brain regions or the electrode placements of consumer EEG devices. The results indicate that using the electrodes placed on the frontal lobe is the best option. It maintains the high classification accuracy, with only around 3-5\% drop in performance compared to the full-head sensor system. The outcome is consistent with previous research \cite{wilaiprasitporn2019affective} which employs data during affective recognition tasks and CNN+LSTM networks. Additionally, the classification performance of our network shows the capability to work with only 14/8 electrodes, as available on the Emotive EPOC+/OpenBCI devices. The performance is up to 82\% CRR using EEG-BBNet[RHO].

\section{Conclusion}
\label{sec:conclusion} 
We introduced EEG-BBNet, a novel brain biometric technique admixing the CNN and GCNN to construct a hybrid framework for personal identification. The benefits of automatic feature extraction from CNN and learning channel-to-channel relationships from GCNN are combined. 
On a benchmark dataset, the model performance is evaluated both within and across sessions using three BCI tasks. The proposed model outperforms previous research and achieved the highest performance on the within-session ERP task. In addition, the results reveal that the EEG frontal lobe is the most recommended region for channel reduction. Our algorithm is also more adaptive to different tasks and sessions than other baselines. However, its utilization in an unseen session requires further research.

\bibliography{References}
\bibliographystyle{IEEEtran}

\end{document}